\documentclass[pdflatex,sn-jnl]{sn-jnl}

\usepackage{url} 
\usepackage{lineno}
\usepackage{wrapfig}
\usepackage{graphicx}
\usepackage{siunitx}
\usepackage{amsmath}
\usepackage{amssymb}
\usepackage{subcaption}
\usepackage{paralist}
\usepackage{newunicodechar}
\usepackage{tabularx} 
\graphicspath{{./figures/}}

\jyear{2022}%

\newcommand{\Xmax}{X_{\textrm{max}}}

\newcommand*\diff{\mathop{}\!\mathrm{d}}
\newcommand*\Diff[1]{\mathop{}\!\mathrm{d^#1}}

\definecolor{color1}{rgb}{0.01176, 0.1015686, 0.3686}
\definecolor{color2}{rgb}{0, 0.4666, 0.7137}
\definecolor{color3}{rgb}{0, 0.70588, 0.8470}
\definecolor{color4}{rgb}{0.5647, 0.8784, 0.93725}
\definecolor{color5}{rgb}{0.7921, 0.9412, 0.9725}

\newenvironment{conditions*}
  {\par\vspace{\abovedisplayskip}\noindent
   \tabularx{\columnwidth}{>{$}l<{$} @{${}={}$} >{\raggedright\arraybackslash}X}}
  {\endtabularx\par\vspace{\belowdisplayskip}}

\usepackage{circledsteps}
\newcommand{\figcirc}[1]{\Circled[fill color=orange,outer color=black,inner color=black]{\textbf{#1}}}

\usepackage{enumitem}
\newlist{figenum}{enumerate}{1}
\setlist[figenum]{font=\color{orange}\bfseries}

\DeclareRobustCommand{\okina}{%
  \raisebox{\dimexpr\fontcharht\font`A-\height}{%
    \scalebox{0.8}{`}%
  }%
}
\newunicodechar{ʻ}{\okina}

\begin{document}

\title[Passive bistatic subsurface radar via the Askaryan effect]{Passive bistatic radar probes of the subsurface on airless bodies using high energy cosmic rays via the Askaryan effect}


\author*[1]{ \fnm{Remy} \sur{Prechelt}}\email{prechelt@hawaii.edu}
\author[2]{ \fnm{Emily} \sur{Costello}}
\author[3]{ \fnm{Rebecca} \sur{Ghent}}
\author[1]{ \fnm{Peter} \sur{Gorham}}
\author[2]{ \fnm{Paul} \sur{Lucey}}
\author[4]{ \fnm{Andrew} \sur{Romero-Wolf}}
\author[1]{ \fnm{Gary} \sur{Varner}}

\affil*[1]{\orgdiv{Department of Physics and Astronomy}, \orgname{University of Hawaiʻi M\=anoa}, \orgaddress{\street{Correa Road}, \city{Honolulu}, \postcode{96822}, \state{Hawaiʻi}, \country{USA}}}

\affil[2]{\orgdiv{Hawaiʻi Institute of Geophysics and Planetology}, \orgname{University of Hawaiʻi M\=anoa}, \orgaddress{\street{East-West Road}, \city{Honolulu}, \postcode{96822}, \state{Hawaiʻi}, \country{USA}}}

\affil[3]{\orgname{Planetary Science Institute}, \orgaddress{\street{E Fort Lowell Road}, \city{Tucson}, \postcode{85719}, \state{Arizona}, \country{USA}}}

\affil[4]{\orgdiv{Jet Propulsion Laboratory}, \orgname{California Institute of Technology}, \orgaddress{\street{Oak Grove Drive}, \city{Pasadena}, \postcode{91109}, \state{California}, \country{USA}}}


\abstract{ We present a new technique to perform passive bistatic subsurface
  radar probes on airless planetary bodies. This technique uses the naturally
  occurring radio impulses generated when high-energy cosmic rays impact the
  body's surface. As in traditional radar sounding, the downward-beamed radio
  emission from each \textit{individual} cosmic ray impact will reflect off
  subsurface dielectric contrasts and propagate back up to the surface to be
  detected. We refer to this technique as \textit{Askaryan radar} after the
  fundamental physics process, the Askaryan effect, that produces this radio
  emission. This technique can be performed from an orbiting satellite, or from a
  surface lander, but since the radio emission is generated beneath the surface,
  an Askaryan radar can completely bypass the effects of surface clutter
  and backscatter typically associated with surface-penetrating radar. We
  present the background theory of Askaryan subsurface radar and show results
  from both finite-difference time-domain (FDTD) and Monte Carlo simulations
  that confirm that this technique is a promising planetary radar sounding
  method, producing detectable signals for realistic planetary science
  applications.}




\maketitle

\section{Introduction}\label{sec:introduction}

Ground penetrating radar (GPR) has been a powerful and unique tool for
performing remote measurements of the subsurface structure and composition of
planetary bodies for nearly five
decades~\cite{1974_Apollo_Lunar_Sounding_Radar}, with previous, current and
future targets including the Moon~\cite{1974_Apollo_Lunar_Sounding_Radar,
  2000_LRS_Selene,2011_Insight_Lunar_Processes_MiniRF,2020_Change5_Lunar_Radar},
Mars~\cite{2007_SHARAD_MRO,2015_MARSIS_Radar,2020_RIMFAX_Mars},
Europa~\cite{2015_REASON}, Ganymede~\cite{2013_JUICE}, and the comet
67P/Churyumov-Gerasimenko~\cite{2007_CONSERT_Comet_Radar}. Ground penetrating
radar soundings have been performed from the body's surface, with a lander or
rover, and from orbit, with a range of radar frequencies and bandwidths tailored
to meet the needs of the particular
investigation~\cite{2016_Radar_SolarSystem_Review}. The fundamental principle of
many of these probes has been the same: an electromagnetic wave is transmitted
into the subsurface from the spacecraft and propagates down into the planetary
body; the radar ``echoes'', reflections from subsurface dielectric contrasts in
the volume are observed by a radio receiver and encode information about the
composition, structure, and dielectric properties of the
subsurface~\cite{2001_Knight_GPR_Environmental_Review,2015_Radar_Planetary_Science_Overview}.



In this work, we present an innovative new technique for performing completely
passive bistatic radar probes of the subsurface on airless or nearly airless
planetary bodies using the naturally occurring broadband radio impulses
generated when high-energy cosmic rays impact the surface. We refer to this
technique as \textit{Askaryan subsurface radar}, named for the fundamental
physics process, the Askaryan effect, that generates this impulsive radio
emission. This technique can be done from orbit, or from the surface, and has
the advantage that the highly-impulsive radar signals are generated
\textit{beneath} the surface, completely bypassing the surface clutter and
backscatter effects typically associated with orbital radar sounding. This
technique has the potential to provide high-horizontal and depth resolution,
commensurate with that of a typical lander or rover mission, with the total area
coverage of orbital radar sounding instruments, while using only passive radio
instrumentation. 

Section~\ref{sec:askaryan-subsurface-radar} finishes the introduction to this
technique, and Section~\ref{sec:data-methods} presents the background theory and
simulation methodology that we use to validate this concept, including a
discussion of the cosmic ray flux in Section~\ref{sec:cosmic-ray-flux}, the
theory of high energy particle cascades in dense media in
Section~\ref{sec:particle-cascades}, and the origin of the radio impulses via
the Askaryan effect in Section~\ref{sec:askaryan-effect}.
Section~\ref{sec:results} presents the results of a pair of simulations of this
technique using both finite-difference time-domain (FDTD) and Monte Carlo (MC)
techniques in Section~\ref{sec:fdtd-results} and
Section~\ref{sec:monte-carlo-results}, respectively, and discusses an immediate application of this technique to the detection of water ice in the permanently shadowed regions at the lunar poles.
\subsection{Askaryan Subsurface Radar}
\label{sec:askaryan-subsurface-radar}

\begin{figure}[htbp]
\centerline{\includegraphics[width=1\textwidth]{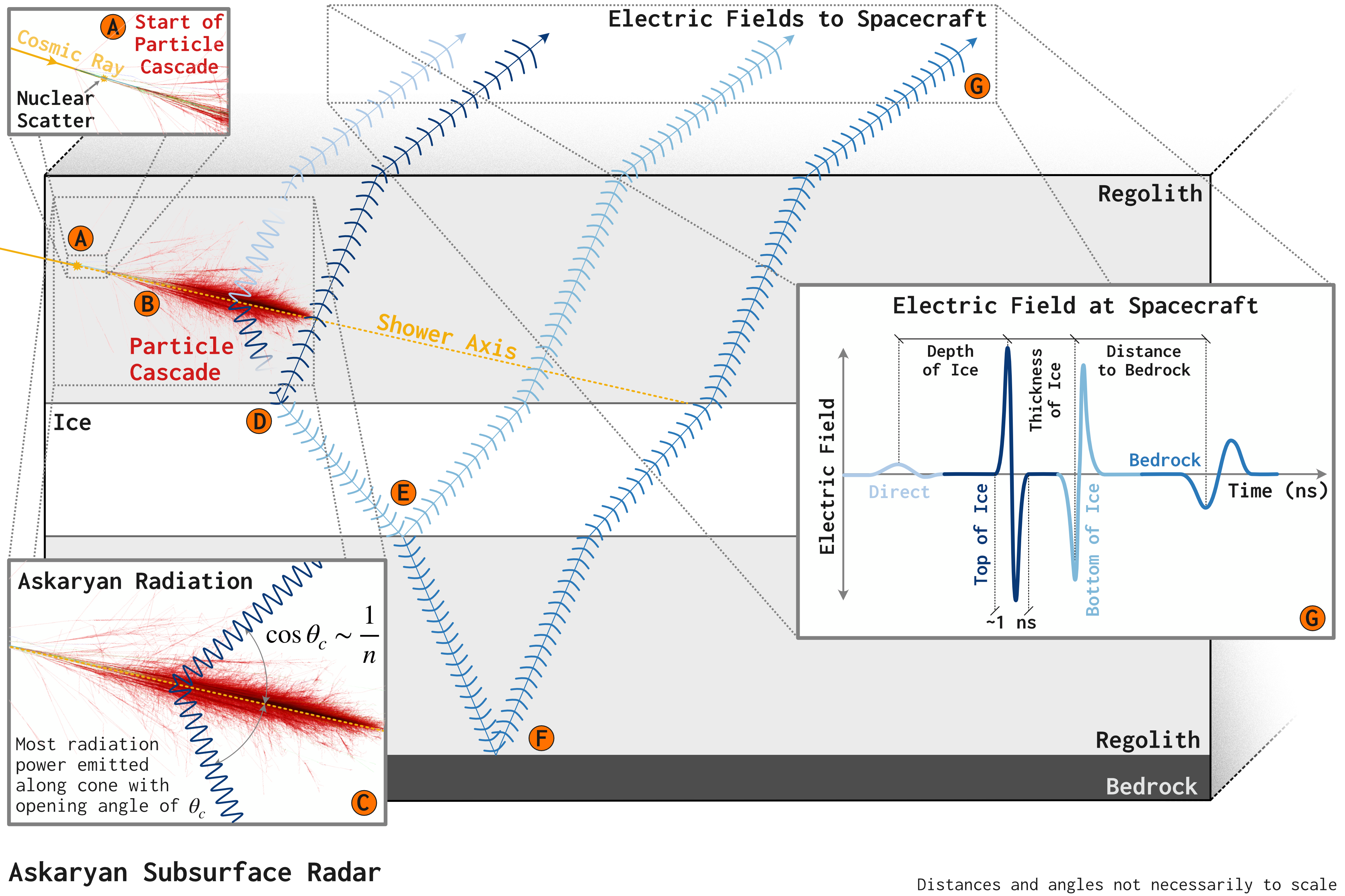}}
\caption[]{\label{fig:askaryan-geometry} A diagram illustrating the
  new \textit{Askaryan subsurface radar} technique proposed in this
  work, for an example application of detecting a coherent ice layer buried in
  regolith. Each orange marker has a corresponding description in the
  text of Section~\ref{sec:askaryan-subsurface-radar}.}
\end{figure}

A diagram demonstrating the key features of the \textit{Askaryan
  radar} technique is shown in Figure~\ref{fig:askaryan-geometry}, for
an example application of detecting a buried ice layer in the lunar
regolith as might potentially exist in the permanently shadowed
regions at the lunar
poles~\cite{2021_Costello_SecondaryImpactBurial_IceDepth}. We
introduce the core concepts of this technique, starting with the
initial cosmic ray impact and ending with the detection of the radio
emission by a radar receiver; each orange marker in the text
corresponds to an identical marker in
Figure~\ref{fig:askaryan-geometry}.

\begin{figenum}[label=\figcirc{\Alph*}]
  \item Planetary bodies in the solar system that lack a dense atmosphere are
  continually bombarded by both solar, galactic, and (likely) extra-galactic
  cosmic rays. When these cosmic rays impact the surface, they almost
  immediately initiate a high energy particle cascade.

  \item This particle cascade, starting with the nuclei from the initial
  collision with the cosmic ray which goes on to collide with other nuclei (and
  so on), develops in length and particle number until it potentially contains
  billions of charged particles (depending upon the cosmic ray energy). In
  high energy physics, these particle cascades are known as \textit{"showers"} and have been a mainstay of particle physics for nearly a century~\cite{PDG2021}.

  \item These high energy particle cascades emit 100\% linearly polarized
  wide-bandwidth coherent radio impulses as they develop in the subsurface; this
  emission process is known as the Askaryan effect. The source of this emission is the coherent radio Cherenkov radiation from the 20\%-25\% compact
  negative charge excess that forms along the front of all high energy particle
  cascades in dense media.



\item\figcirc{E} \figcirc{F} The Askaryan emission from the cascade
  propagates down into the subsurface, with unique reflections
  generated by each dielectric contrast in the volume (in this example,
  the regolith-ice, ice-regolith, and regolith-bedrock interfaces)
  just as in traditional radar sounding.
\addtocounter{figenumi}{2}
\item An example of the signals that might be observed by a spacecraft
  for the subsurface strata of Figure~\ref{fig:askaryan-geometry}, in
  addition to the direct signal observed \textit{without} reflection,
  is shown in inset \figcirc{G} and encode valuable information about
  the structure and composition of the subsurface.


\end{figenum}

This technique has been previously performed \textit{on Earth} with the
Antarctic Impulsive Transient Antenna (ANITA). ANITA, a long-duration
balloon-borne cosmic ray and neutrino observatory, has detected
\(\mathcal{O}(60)\) ultrahigh energy cosmic rays via their impulsive radio
emission after it was\textit{reflected off} the Antarctic surface (a completely
analogous process to the detection of lunar ice described
above)~\cite{2010_Hoover_CosmicRay_Discovery}. While these detections are not
scientifically interesting on Earth due to existing high resolution measurements
of both the cosmic ray flux~\cite{2017_Auger_Combined_Fit}, and the Antarctic
surface from airborne and orbital radar
instruments~\cite{2009_Antarctic_Ice_Sheet}, they do provide important
validation of, and heritage to, this technique. Therefore, the question posed by
this work is as follows: is the detection rate and sensitivity of an Askaryan
radar mission scientifically compelling for use in planetary science applications?

\section{Background theory \& simulation methods}
\label{sec:data-methods}


The underlying physics of cosmic ray-induced cascades, and their associated
radio emission via the Askaryan effect, is extremely well studied with extensive
terrestrial measurements by dozens of observatories over the last
century~\cite{PDG2021}. In the following sections, we present a summary of the
theory of particle cascades as it relates to our concept and discuss the data
and models needed to perform accurate simulations of this technique.

\subsection{Background theory}

In the following sections, we present the background theory of this technique,
including the high energy cosmic ray flux, the physics of particle cascades, and
the generation of the radio emission via the Askaryan effect.

\subsubsection{Flux of high energy cosmic rays}
\label{sec:cosmic-ray-flux}

\begin{figure}[hb]
  \centering
\includegraphics[width=0.7\textwidth]{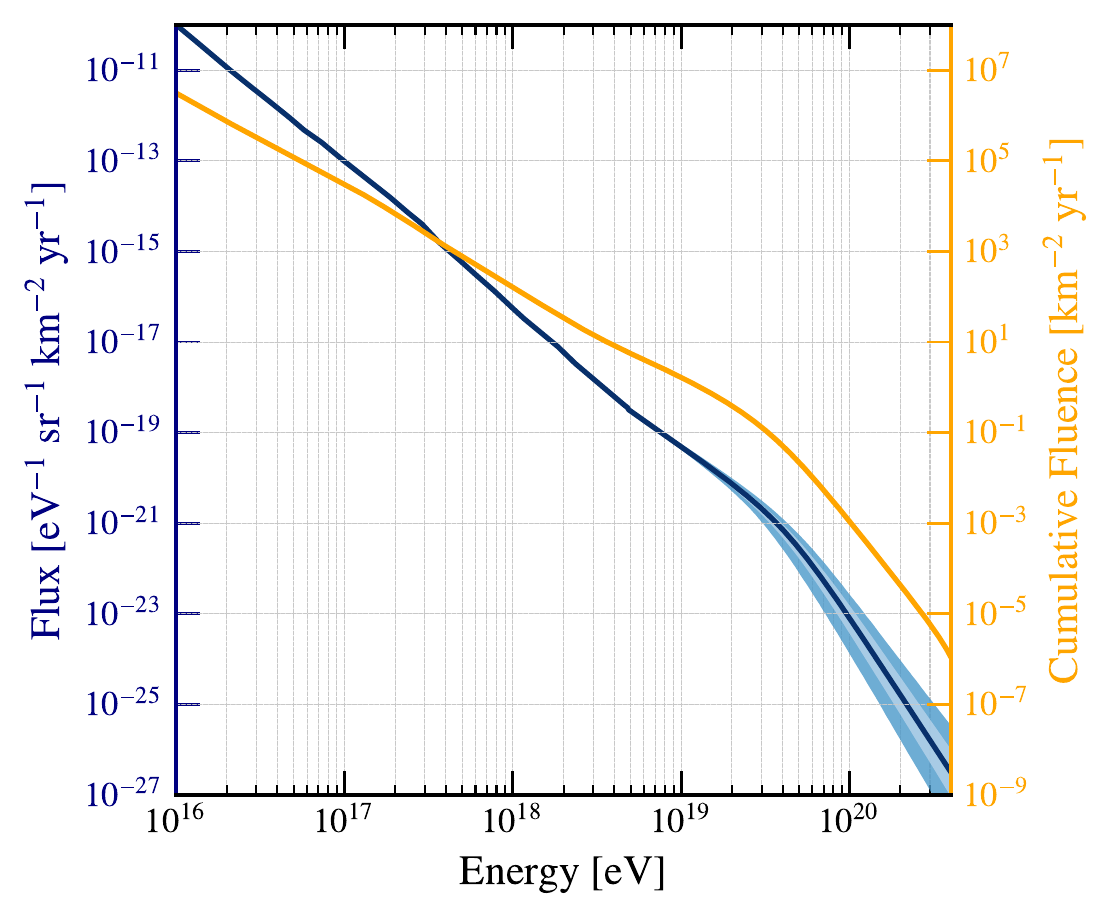}
\caption{ The all-particle flux of high energy and ultrahigh energy cosmic rays
  measured by the Pierre Auger observatory from
  \SI{10}{PeV} to \SI{500}{EeV} is shown in blue (left axis) along with the
  \(1\sigma\) (light blue) and \(2\sigma\) (dark blue) uncertainty region for
  the measurement, which starts to become significant above \SI{5}{EeV}. The inverse
  cumulative fluence, the total fluence of particles \textit{above} a given
  energy (\SI{}{km^{-2}.yr^{-1}}), is shown in orange (right
  axis)~\cite{2017_Verzi_EnergySpectrum_UHECR}.\label{fig:cr-flux}}
\vspace{-2mm}
\end{figure}

The flux of high energy cosmic rays is a power-law-\textit{like} flux,
\(E^{\gamma(E)}\), with exponent, \(\gamma(E)\), varying from -2.7 to -3.3, in
the energy range applicable to this technique. The Askaryan radar technique is
currently most amenable to cosmic rays with energies above roughly \SI{10}{PeV}
(\(\SI{e16}{eV}\)) due to thermal noise constraints on the radio detection, and
extending up to the highest observed cosmic rays, so-called ultrahigh-energy
cosmic rays (UHECRs), at \({\sim}\SI{500}{EeV}\) (\(\SI{5e20}{eV}\)). The flux
of high energy and ultrahigh energy cosmic rays over this energy range is shown in Figure~\ref{fig:cr-flux}. Above
\SI{100}{PeV}, approximately \({\sim}\SI{4e4}{}\) cosmic rays impact every
square kilometer of a planetary body per year, but due to the steeply falling
cosmic ray flux spectrum (\(\propto E^{\approx -3}\)), lower energy cosmic rays
are the dominant contributor to the overall detection rate given a fixed cosmic
ray energy threshold. Due to their propagation through galactic and
extra-galactic fields, the arrival direction of high energy cosmic rays, which
can likely propagate from sources at distances up to
\(\mathcal{O}(\SI{100}{Mpc})\), are highly isotropic with respect to the surface
of any airless body in the solar system and can be assumed to arrive (stochastically) uniformly in time and in solid angle.



\subsubsection{Cosmic ray cascades in dense media}
\label{sec:particle-cascades}

\begin{figure}
     \centering
     \begin{subfigure}[b]{0.495\textwidth}
       \centering
       \includegraphics[width=\textwidth]{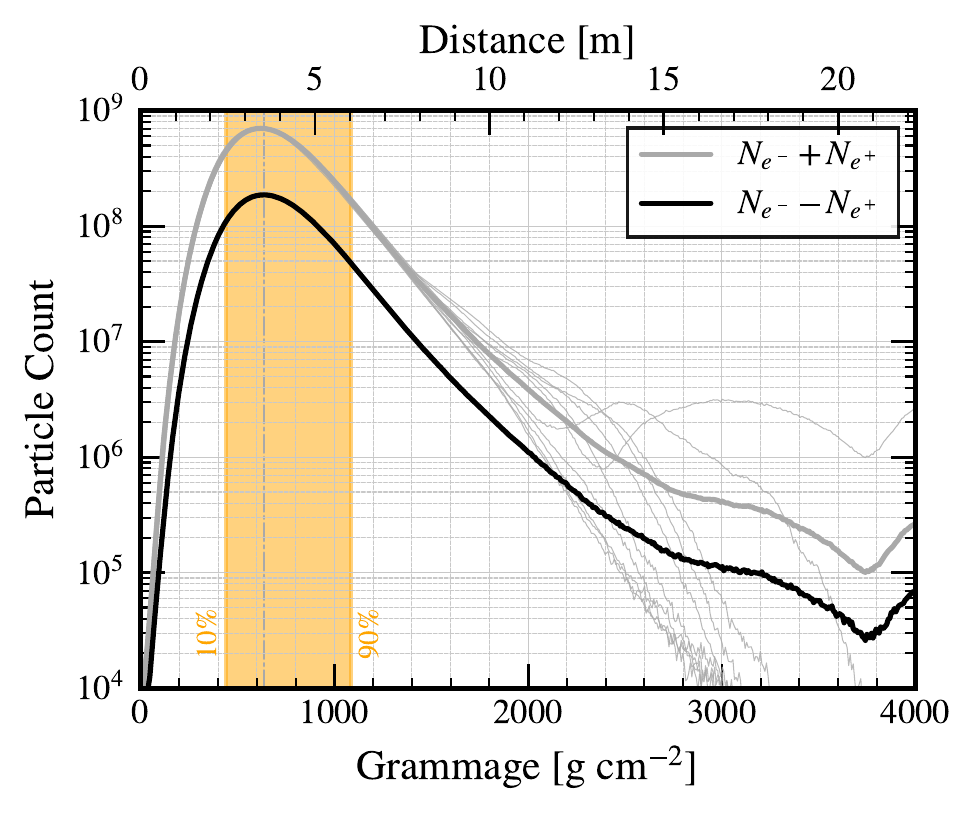}
       \caption{Lunar Regolith}
       \label{fig:regolith-shower-profiles}
     \end{subfigure}%
     \hfill
     \begin{subfigure}[b]{0.495\textwidth}
       \centering
       \includegraphics[width=\textwidth]{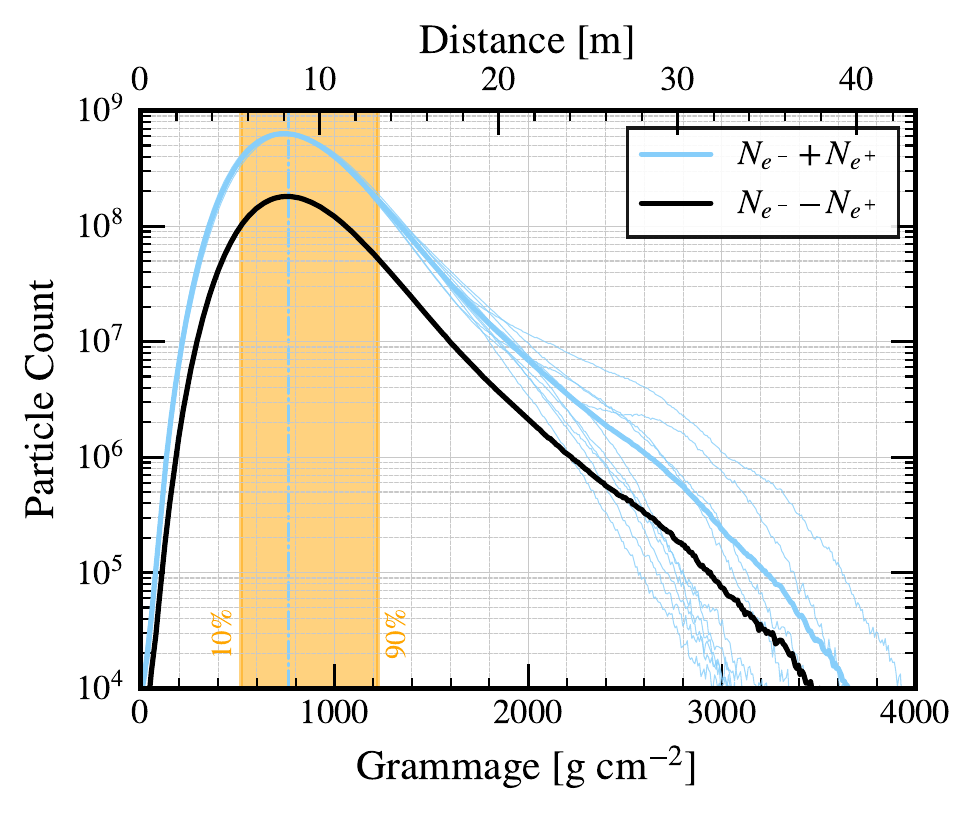}
       \caption{Ice}
       \label{fig:ice-shower-profiles}
     \end{subfigure}%
     \caption{The longitudinal profile of the total electronic charge (electrons
       \textit{and} positrons in grey or blue), and the profile of the negative
       charge excess (electrons \textit{minus} positrons in black) for a sample
       of \SI{1}{EeV} proton-initiated cascades in lunar regolith and in ice.
       The thick grey (blue) and black lines indicate the average profile over)
       ten showers for regolith (ice), while the thin lines show the individual
       profiles from each simulated shower. The orange band indicates the 10\%
       to 90\% cumulative charge length for the average profiles and represents
       the ``bulk'' of the shower development. The particle cascades were
       simulated with the TIERRAS code~\cite{2010_TIERRAS}, a modification of
       the ZHAireS code for use in dense media~\cite{2012_MonteCarlo_AtmosphericShowers_ZHAireS}. For all simulations, lunar
       regolith is simulated with a nominal density of
       \(\rho=\SI{1.8}{g.cm^{-3}}\) and refractive index \(n=1.78\), and ice is
       simulated with \(\rho=\SI{0.924}{g.cm^{-3}}\) and
       \(n=1.305\)~\label{fig:shower-profiles}}
   \end{figure}

When primary high energy cosmic rays impact the dense surface of a nearly
airless planetary body, they initiate particle cascades via high energy
inelastic collisions with nuclei in the medium. These cascades develop predominantly along the
original direction of the cosmic ray (known as the \textit{shower axis}---see
Figure~\ref{fig:askaryan-geometry}) and grow in size until they reach a maximum
number of particles, at which point energy losses to the medium continually slows
the production of new particles until the shower development stops (as existing
particles in the cascade no longer have sufficient mass-energy to produce new
particles in the shower). The longitudinal profiles, the number of particles in the cascade
projected onto the shower axis, of a population of \SI{1}{EeV} proton-initiated
cosmic ray cascades in the lunar regolith and in ice is shown in
Figure~\ref{fig:regolith-shower-profiles} and
Figure~\ref{fig:ice-shower-profiles}, respectively.

The development of a high energy cascade is driven by the energy of
the cosmic ray primary and the total grammage (density multiplied by
length, measured in \(\SI{}{g.cm^{-2}}\)) experienced by the shower as
it develops. The grammage, usually denominated as \(X\), where the
number of particles in the shower is at a maximum is known as
\textit{shower maximum}, and is denoted as \(\Xmax\). As medium
density decreases, the average length of these showers increases;
showers in regolith-like materials are typically
\(\mathcal{O}(\SI{10}{m})\) long, with \(\Xmax\) occurring at
\({\sim}\SI{4}{m}\), whereas showers in ice or air can be tens or
thousands of meters in length, respectively. However, when expressed in terms of grammage (top axis of Figure~\ref{fig:regolith-shower-profiles} and
Figure~\ref{fig:ice-shower-profiles}), showers in different media are highly comparable with shower maximum typically occurring at \(\SIrange{600}{800}{g.cm^{-2}}\) at \SI{1}{EeV} with \(\Xmax\) increasing as the primary particle energy increases.





   \subsubsection{Development of a negative charge excess in high energy cascades}
\label{sec:charge-excess}


These particle cascades develop a 20\%-25\% negative charge excess as they
develop, as atomic electrons are preferentially upscattered into the shower and
positrons (produced via pair production in the cascade) are annihilated against
this same population of atomic electrons. Therefore, high energy particle
cascades are not electrically neutral and this negative charge excess is
typically centimeters thick and millimeters-wide in dense media like regolith
and is concentrated along the front of the
cascade~\cite{1960_Askaryan_Prediction}. The longitudinal profile of negative
charge excess of the a population of \SI{1}{EeV} proton showers is shown in
the solid black lines of Figure~\ref{fig:regolith-shower-profiles} and
Figure~\ref{fig:ice-shower-profiles}.

\subsubsection{Askaryan Effect}
\label{sec:askaryan-effect}



All charged particles moving faster than the phase speed of light in a medium
(\(c/n\)) emit broadband Cherenkov radiation that is typically observed in the
optical and near-ultraviolet wavelengths (due to the Cherenkov power spectrum increasing
with frequency)~\cite{Cherenkov1934,1960_Tamm_CherenkovRadiation,PDG2021}. When
observed at wavelengths larger than the physical size of this compact charge
excess, typically a few \SI{}{MHz} to a few \SI{}{GHz} depending upon the
medium, the radio-wavelength Cherenkov radiation from each particle in the
cascade is observed \textit{coherently} and the entire charge excess is observed
as a relativistic particle emitting coherent radio
Cherenkov emission~\cite{Saltzberg:2000bk}. The coherent radio and microwave emission from
the negative charge excess in a high energy shower is known as the
\textit{Askaryan effect} and has been experimentally measured in a variety of
media over more than two
decades~\cite{Saltzberg:2000bk,Gorham:2004ny,ANITA:2006nif,Gorham:2017nzv}, and forms the basis for a multiple existing cosmic ray and neutrino observatories on Earth~\cite{2011_ARA,2019_ARIANNA,2006_ANITA_Intro}.

The electric field created by the Askaryan effect is 100\% linearly polarized,
with a polarization vector that is uniquely determined by the cosmic ray
trajectory and observation direction, and the emitted electric fields are
extremely impulsive with typical durations of \(\mathcal{O}(\SI{1}{ns})\)
close to the angle of peak emitted power (the Cherenkov
angle)~\cite{2011_alvarez-muniz:practical_askaryan}. The wide bandwidth and high
linear polarization of the emission from the Askaryan effect makes it a strong
potential candidate for use in surface penetrating radar.

   \begin{figure}[htb]
     \centering
     \begin{subfigure}[b]{0.495\textwidth}
       \centering
       \includegraphics[width=\textwidth]{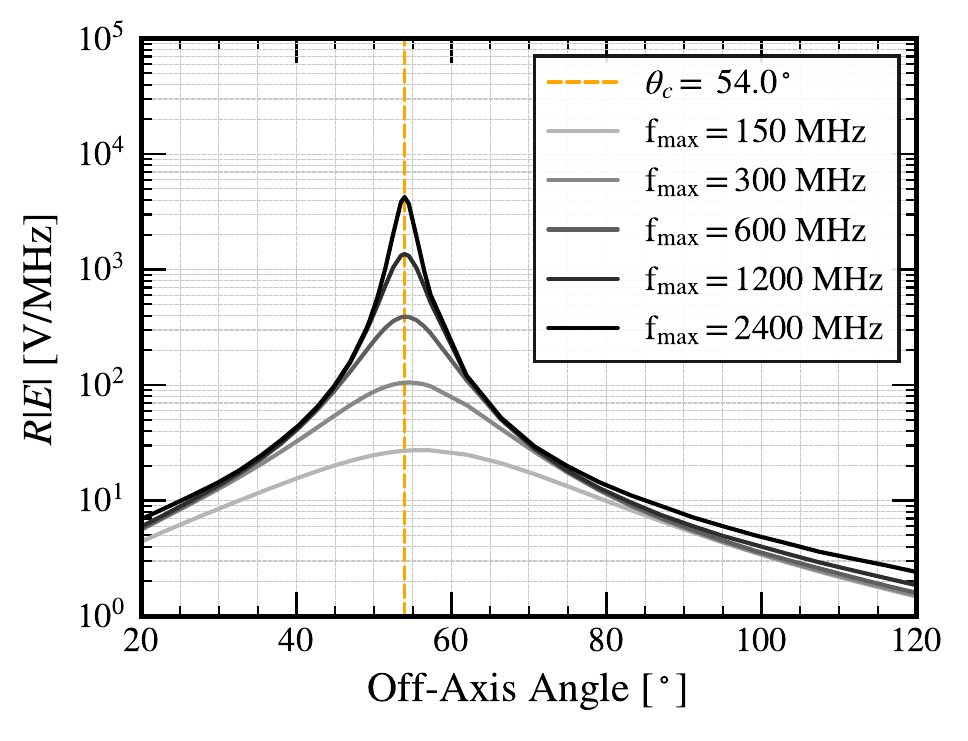}
       \caption{Lunar Regolith}
       \label{fig:regolith-angular-spectrum}
     \end{subfigure}
     \hfill
     \begin{subfigure}[b]{0.495\textwidth}
       \centering
       \includegraphics[width=\textwidth]{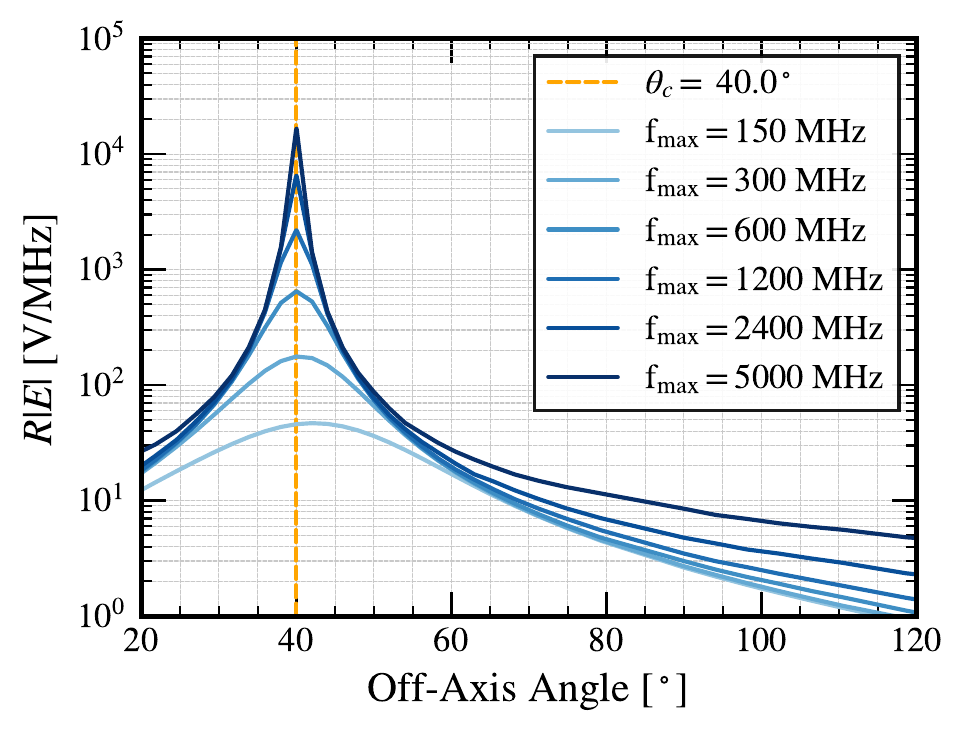}
       \caption{Ice}
       \label{fig:ice-angular-spectrum}%
     \end{subfigure}%

     \begin{subfigure}[b]{0.495\textwidth}
       \centering
       \includegraphics[width=\textwidth]{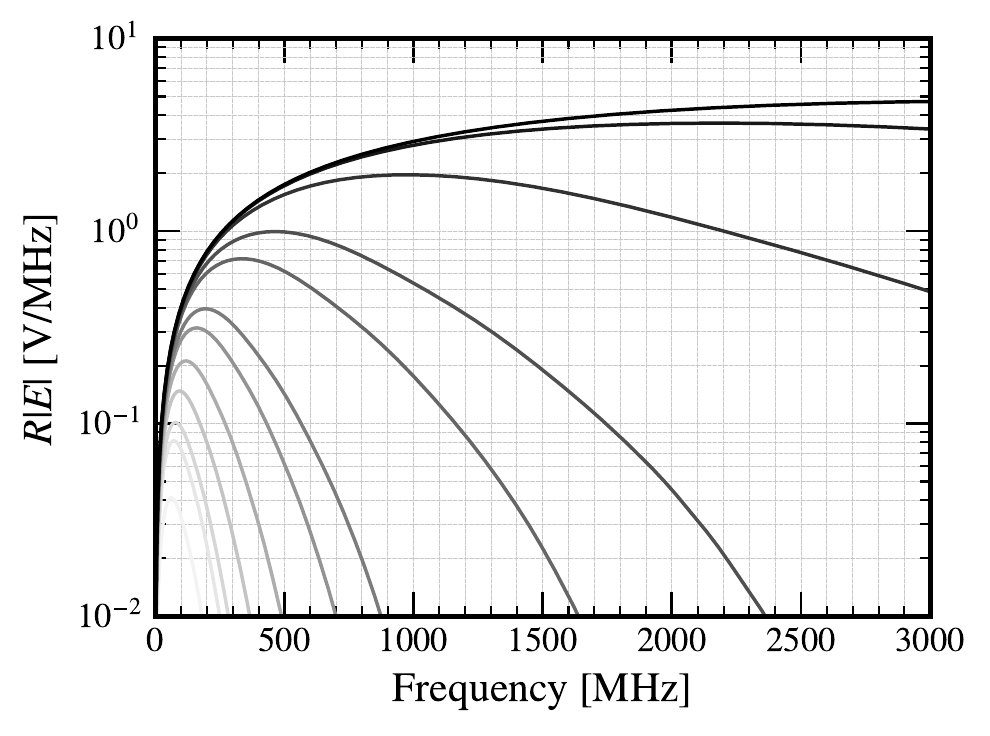}
       \caption{Lunar Regolith}
       \label{fig:regolith-freq-spectra}
     \end{subfigure}%
     \hfill
     \begin{subfigure}[b]{0.495\textwidth}
       \centering
       \includegraphics[width=\textwidth]{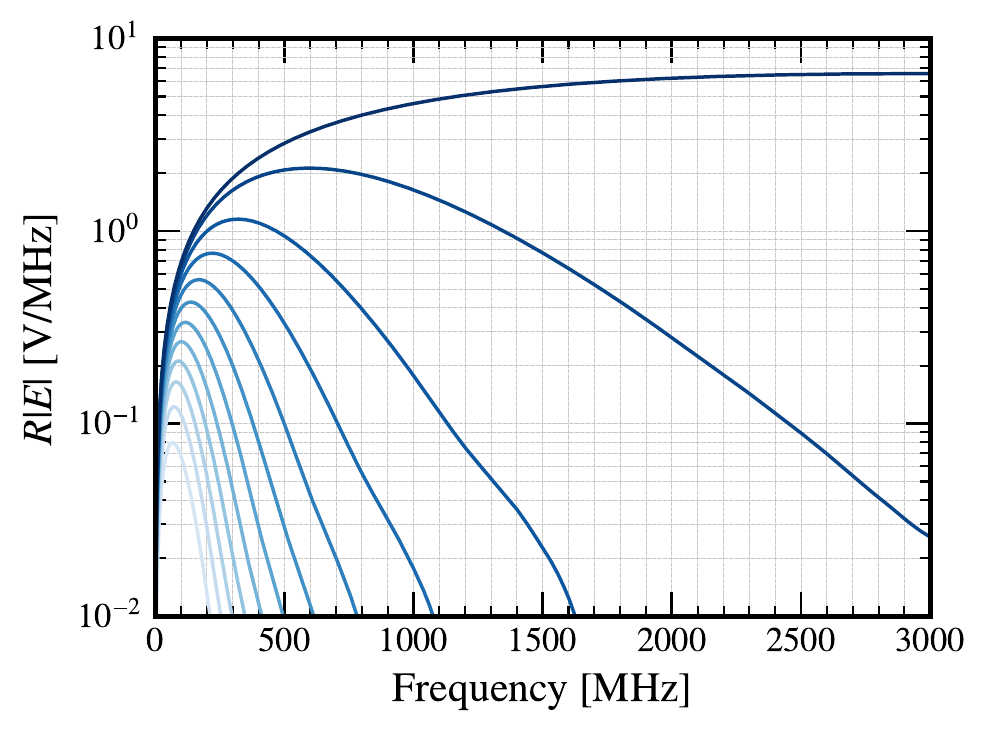}
       \caption{Ice}
       \label{fig:ice-freq-spectra}
     \end{subfigure}%
     \caption{\textit{Top}: The angular spectrum of the electric potential
       spectral density for various upper frequency limits on the
       Askaryan spectrum. The vertical orange line indicates the
       location of the Cherenkov angle in the particular simulation
       configuration.%
       ~\textit{Bottom}: The average electric potential spectral
       density at various off-axis angles for the same cascades in
       regolith and ice, respectively. Near the Cherenkov angle, the
       Askaryan spectrum extends well above \SI{3}{GHz} in both
       media when observed at the Cherenkov angle. For all simulations, lunar regolith is simulated
       with a nominal density of \(\rho=\SI{1.8}{g.cm^{-3}}\) and
       refractive index \(n=1.78\), and ice is simulated with
       \(\rho=\SI{0.924}{g.cm^{-3}}\) and
       \(n=1.305\)~\label{fig:shower-emission}}
   \end{figure}

The total radio-frequency power emitted via the Askaryan effect scales
quadratically with the energy of the incident cosmic ray, so an experiment sensitive to three orders of magnitude of cosmic ray energy (i.e. \SIrange{0.5}{500}{EeV}) will observe six orders of magnitude in radio power.  This power is
emitted with a conical beam pattern aligned with the shower axis (i.e. uniformly
in azimuth around the shower axis but narrowly beamed in polar "off-axis" angle). The average angular spectrum of
the electric potential for the same population of proton-initiated cascades in
regolith and ice is shown in Figure~\ref{fig:regolith-angular-spectrum} and
Figure~\ref{fig:ice-angular-spectrum}. The polar opening angle of this conical
beam is determined by the Cherenkov angle  and is given by
\(\cos\theta \sim 1/n\), where \(n\) is the refractive index of the medium
(inset \figcirc{C} in Figure~\ref{fig:askaryan-geometry}), with a beamwidth
(``thickness'' of the conical beam) that depends on frequency but is nominally a
few degrees for wide bandwidths.

The Askaryan frequency spectrum shows the same characteristic linear rise over
frequency as the underlying Cherenkov emission; this linear rise continues
down to wavelengths such that the cascade is no longer observed coherently by an
observer (when the size of the charge excess becomes comparable to wavelength). This turnover frequency is dependent upon the angle between the
observation direction and the shower axis (the ``off-axis'' angle). The average
frequency spectra of the same population of cascades in regolith and ice is
shown in Figure~\ref{fig:regolith-freq-spectra} and
Figure~\ref{fig:ice-freq-spectra}, respectively. When observed close to the
Cherenkov angle, where the coherence is maximized, Askaryan radiation can be
extremely broadband, extending to \({\gtrsim}\SI{4}{GHz}\) in both ice and
regolith; as the observer moves further \textit{off-cone} (i.e. away from the
Cherenkov angle), the cascade is no longer observed coherently and the upper
limit of the spectral density decreases.

\subsection{Simulation Methods}
\label{sec:methods}

We analyze our Askaryan radar concept using two different simulation methods:
\begin{inparaenum}[(a)]
\item a finite-difference time-domain (FDTD) simulation that is highly
  accurate, capturing the full-wave electromagnetic behavior of the
  Askaryan emission process and subsurface reflection, but is computationally
  intractable for use in a comprehensive study of this technique; and,
  \item a semi-analytical Monte Carlo (MC) simulation that is orders of
  magnitude faster, at the cost of lower fidelity, and is therefore
  computationally tractable for exploring the available phase space.
\end{inparaenum}

We perform these simulations for the simplified geometry of
Figure~\ref{fig:askaryan-geometry}---a subsurface layer of ice buried within an
otherwise uniform volume of regolith---but we vary the depth, thickness, roughness, and
composition of the buried layer to probe the phase space of subsurface structure
and composition detectable by an Askaryan radar for this application. This example geometry is
motivated by the search for subsurface water ice on the Moon that may have been
buried by hydrated impacts and preserved under several meters of regolith in
the permanently shadowed regions at the lunar poles for geologic
time~\cite{2021_Costello_SecondaryImpactBurial_IceDepth}. 


\subsubsection{Finite-Difference Time-Domain Simulations}
\label{sec:fdtd}

The finite-difference time-domain (FDTD) algorithm is a technique for performing
high-fidelity computational electrodynamics simulation for arbitrary geometries
and sources. The FDTD method discretizes Maxwell's equations on a
spatio-temporal grid and directly evolves the electric and magnetic field
components in each grid cell over time enabling accurate simulations of the
generation of Askaryan emission from a relativistic shower, and the subsequent
reflection off realistic subsurface
layers~\cite{1966_FDTD_Yee}. 
While highly accurate, FDTD simulations are extremely computationally
expensive and are therefore limited to relatively small simulation
domains without becoming intractable. In our work, we
simulate the volume of the subsurface reflector over several Fresnel zones,
thus typically $50 \times 10 \times 15$~m$^3$ (length by width by depth), and an additional volume above the
surface, $50 \times 10 \times 20$~m$^3$ (length by width by height) to ensure 
the that radiation has transitioned to the Fraunhofer zone
where we can propagate the fields, using analytical
transforms, to the electric field detected by the
spacecraft in the far field.


As there is no currently available FDTD code that can directly simulate the
evolution of high energy cosmic ray induced particle cascades, we first simulate
proton-initiated cascades in the lunar regolith using
TIERRAS~\cite{2010_TIERRAS}---a well-validated code for the simulation of cascades in dense media---and use the longitudinal profile of the charge excess
(Figure~\ref{fig:regolith-shower-profiles} and
Figure~\ref{fig:ice-shower-profiles}) to construct an equivalent relativistic
time-varying current source in the commercial XFDTD software that reproduces the
simulated charge excess profiles. We use the commercial FDTD software XFdtd to perform each simulation~\cite{xfdtd}. 
The Askaryan radiation is automatically generated by the discretized
form of Maxwell's equations from the inserted current profile,
propagates down into the subsurface, and reflects off a rough
surface based on the ChangE-3 Yutu rover data for a 60~m long
track, which implies an RMS height deviation of about \SI{0.5}{m} over the \SI{60}{m} range~\cite{2014_LPR_Change3,2017_3D_Geologic_Model_Change3}. To translate this measurement
to the $\sim 2$~m typical diameter of the Fresnel zone for the Askaryan pulse reflection, 
we simulate a \SI{50}{m} diameter 2D self-affine (eg. fractal) surface~\cite{Hurst} 
with characteristics based on the measured rover track RMS and power spectral density, 
and then sample the sub-region RMS over patches with the average
size of the Fresnel zone, with a plane-slope removed.

\subsubsection{Monte Carlo Simulation Method}
\label{sec:monte-carlo}

As this technique relies on cosmic ray impacts that occur stochastically over
time, area, and solid angle, an Askaryan radar instrument must continually
observe a given surface waiting for a cosmic ray impact to illuminate the
subsurface layer from an unknown direction. The rate of subsurface radar detections (referred to as \textit{``events''} in
the following text) made by an Askaryan radar depends on the cosmic ray flux,
the physical properties of the subsurface, the geometry of the observation, and
the sensitivity of the spacecraft. Given a distribution of subsurface
reflectors embedded in a subsurface volume, the \textit{average} detection rate,
\(\diff\bar{N}/\diff t\), can be approximately expressed as: %

\begin{minipage}{\textwidth}
\begin{equation}
  \label{eq:monte-carlo-evaluation}
\begin{aligned}
\frac{\diff\bar{N}}{\diff t} =
&\begingroup
\color{black}
\underbrace{\color{black} \int_{E_{\mathrm{min}}}^{\SI{e21}{eV}} \diff E\, \mathcal{F}(E)}_{\text{Cosmic ray flux}}
\endgroup
\begingroup
\color{color1}
\underbrace{\color{black} \int_{\Sigma}\diff \vec{\sigma} \int_{\Omega}\diff \vec{\omega}}_\text{Geometric acceptance}
\endgroup\\
&\begingroup
\color{color2}
\underbrace{\color{black} \int_{0}^{R_{\text{moon}}} \int_{0}^{\infty} \diff z\, \diff\Delta n\, \tilde{P}(z, \Delta n)}_{\text{Depth distribution of reflectors}}
\endgroup\\
&\begingroup
\color{color3}
\underbrace{\color{black} \int \diff X\, \tilde{P}(\vec{X}_{\mathrm{max}} \vert E, \vec{\omega}, \vec{\sigma})}_{\text{Location of } \vec{X}_{\mathrm{max}}}
\endgroup\\
&\begingroup
\color{color4}
\underbrace{\color{black} \int_{\mathcal{E}_{\mathrm{min}}}^{\infty} \diff\mathcal{E}_f\, \tilde{P}(\mathcal{E}_f \vert E, \vec{\sigma},\vec{\omega}, \vec{x}(t), \vec{X}_{\mathrm{max}}, \Delta n, z)}_\text{Reflected Askaryan radiation at spacecraft}
\endgroup
\end{aligned}
\end{equation}
where,
\begin{conditions*}
  \bar{N} & the \textit{average} number of subsurface radar detections (events) made by the spacecraft, \\
  \diff\vec{\sigma} & an element of differential area on the planetary body, \\
  \Sigma & the total area of the body visible from the spacecraft at time \(t\), \\
  \diff\vec{\omega} & an element of differential solid angle centered at \(\diff\vec{\sigma}\), \\
  \Omega & the total solid angle of the cosmic ray flux visible from \(\diff\vec{\sigma}\), \\
  E & the energy of the incident cosmic ray, \\
  E_{\mathrm{min}} & the minimum cosmic ray energy the spacecraft is able to detect; this depends directly on \(\mathcal{E}_{\mathrm{min}}\), \\
  \mathcal{F}(E) & the incident isotropic cosmic ray flux (\SI{}{km^{-2}.s^{-1}.sr^{-1}.eV^{-1}}), \\
  z & depth below the surface of the moon, \\
  \Delta n & the dielectric contrast between two subsequent differential depths, \\
  \tilde{P}(z, \Delta n) & the probability density of finding a dielectric contrast $\Delta n$ at a depth $z$, \\
  X & the grammage experienced by the cosmic ray shower as it develops along the shower axis, \\
  \tilde{P}(\vec{X}_{\mathrm{max}} \vert \ldots) & the probability density of
  shower maximum occurring at a grammage \(\vec{X}_{\text{max}}\) given the parameters of the cosmic ray, \\
  \mathcal{E}_{f} & the electric field from the subsurface reflector \textit{at the spacecraft} , \\
  \mathcal{E}_{\mathrm{min}} & the \textit{minimum} electric field \textit{detectable by the spacecraft}, \\
  \tilde{P}(\mathcal{E}_{f} \vert \ldots) & the probability density of observing an electric field of \(\mathcal{E}_{f}\) \textit{at the spacecraft} given the parameters of the event.
\end{conditions*}
\end{minipage}

The complexity of propagating the Askaryan emission from the shower, to each
subsurface layer, and back up to spacecraft to be detected is included in the
function \(\tilde{P}(\mathcal{E}_{f}\vert \ldots )\).

While the FDTD method provides the highest fidelity for \textit{individual}
cosmic ray impacts, it is computationally intractable to use FDTD simulations to
evaluate \(\tilde{P}(\mathcal{E}_{f} \vert \ldots)\) in the high-dimensional
integral of Equation~\ref{eq:monte-carlo-evaluation}. Therefore, to evaluate
Equation~\ref{eq:monte-carlo-evaluation}, we use semi-analytical models in a
Monte Carlo (MC) integration scheme. This is many orders of magnitude faster
than performing full FDTD simulations for each individual cosmic ray impact and
allows for a study of the phase space of subsurface structures and compositions that could be
detected by an Askaryan radar instrument. 


The methodology for a single Monte Carlo trial, used to
evaluate Equation~\ref{eq:monte-carlo-evaluation}, is as follows:
\begin{enumerate}
  \item We start by choosing an impact point uniformly on the surface of the
        planetary body within view of the spacecraft and choose a cosmic ray
        direction uniformly in solid angle.
  \item We sample a grammage for shower maximum from the measured distribution
        of \(\langle X_{\mathrm{max}}\rangle\) and
        \(\langle \sigma_{X_{\mathrm{max}}}\rangle\) (the fluctation in
        \(X_{\mathrm{max}}\) on a shower-by-shower basis) as a function of
        cosmic ray energy as measured by the Pierre Auger
        Observatory~\cite{2017_Combined_Fit_Spectrum_Composition}.
  \item We then use a pre-calculated library of cosmic-ray regolith cascades
        simulated using the TIERRAS code to estimate the longitudinal charge
        profile of the shower along the shower axis~\cite{2010_TIERRAS}.
  \item For each allowable propagation path between \(X_{\mathrm{max}}\) and the
        observer (spacecraft), we use the semi-analytical method of
        Reference~\cite{2011_alvarez-muniz:practical_askaryan}~to calculate the
        electric field waveform that is emitted in the direction of the observer
        (along both the direct and reflected propagation paths to the
        spacecraft).
  \item We propagate these emitted electric fields down to each subsurface layer,
        apply the appropriate reflection coefficients, propagate them back up to
        the surface, and refract each signal out to the spacecraft (using a
        geometrical optics propagation model), including the Fresnel
        coefficients at all relevant surface and subsurface boundaries, the
        attenuation due to propagation in regolith, and the divergence due to
        the near field refraction at the surface (as the particle cascade acts
        as a near-field line source with respect to the surface refraction
        point).
  \item The direct and reflected electric fields at the spacecraft are then
        compared with a spacecraft instrument model to determine whether this
        specific Monte Carlo trial would have triggered the observatory and been
        detected.
\end{enumerate}

Finally, the collected sample of Monte Carlo trials are weighted by
the cosmic ray flux to
calculate the rate of detected events. For this simulation, we
directly evaluate the differential flux of ultrahigh energy cosmic
rays, \(\Diff4 N / (\diff E\,\diff A\,\diff\Omega\,\diff t)\), as measured by the Pierre Auger
Observatory~\cite{2017_Combined_Fit_Spectrum_Composition}.

For the specific lunar simulation of Figure~\ref{fig:askaryan-geometry}, we
model the lunar regolith as a uniform volume with density \SI{1.8}{g.cm^{-3}},
uniform radio refractive index of 1.78, and a radio loss tangent of \SI{6.5e-4}{}
in the cold permanently shadowed regions at the lunar poles. The subsurface ice
is simulated with varying thicknesses, depths, and purities with 100\% pure ice
assumed to have a density of \SI{0.924}{g.cm^{-3}} and a refractive index of
1.305 (chosen to be consistent with the ice deposition models
of~\cite{2021_Costello_SecondaryImpactBurial_IceDepth}); other purities are
modeled as a combination of regolith and ice with properties calculated via a
composition-fraction weighted sum of the logarithm of the refractive indices of ice and regolith, respectively.




\section{Results}
\label{sec:results}

In this section, we present the results of our finite-difference time-domain and
Monte Carlo studies of the Askaryan radar technique in the example application
of searching for buried ice at the lunar poles.

\subsection{Finite-Difference Time-Domain Results}
\label{sec:fdtd-results}

\begin{figure}
     \centering

     \begin{subfigure}[b]{0.95\textwidth}
       \centering
       \includegraphics[width=\textwidth]{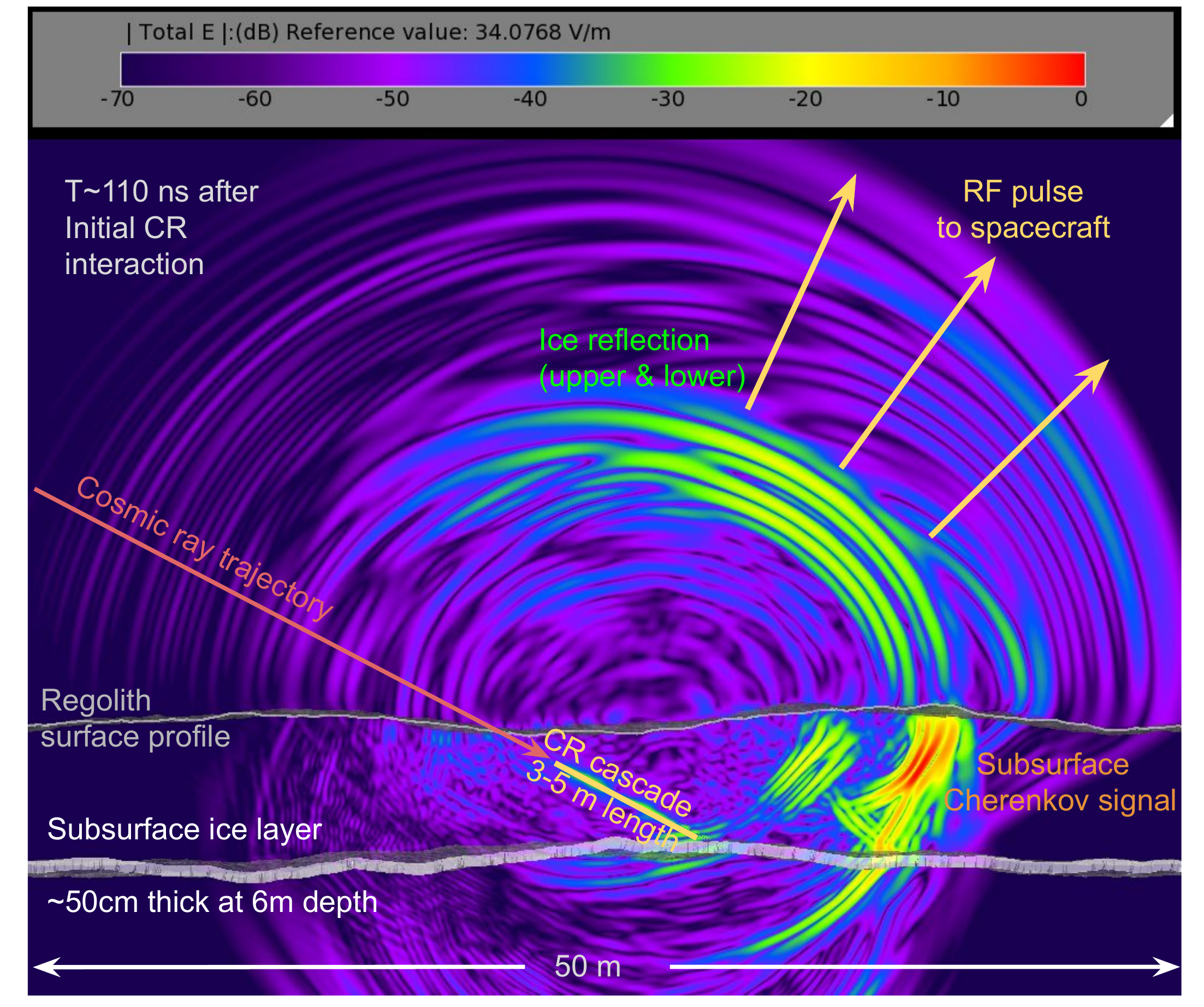}
       \caption{}
       \label{fig:fdtd-slice}
     \end{subfigure}%

     \begin{subfigure}[b]{0.95\textwidth}
       \centering
       \includegraphics[width=\textwidth]{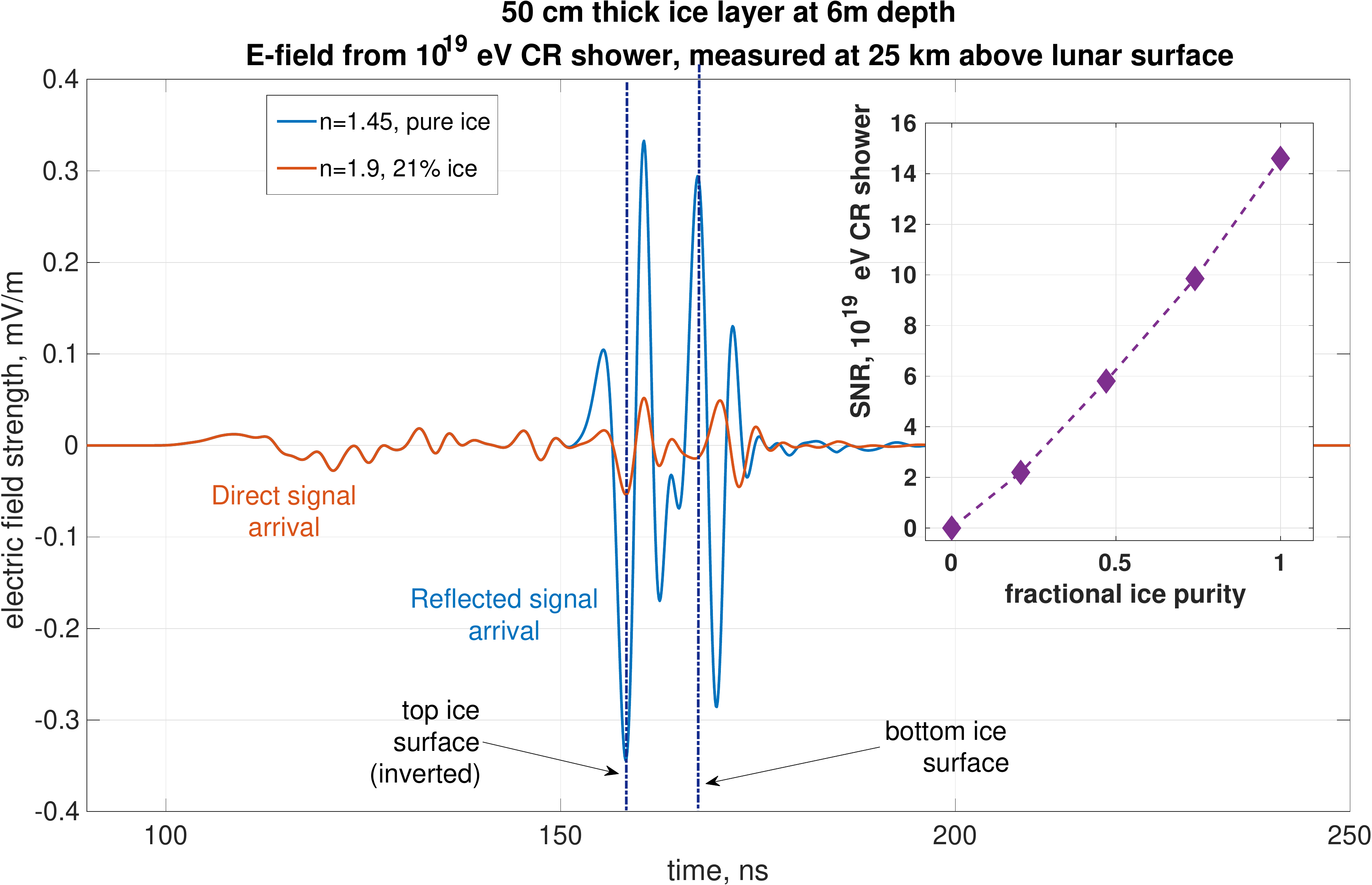}
       \caption{}
       \label{fig:fdtd-waveforms}
     \end{subfigure}
     \caption{\textit{a)} A vertical slice of the electric field
       magnitude from a 3D FDTD simulation of the Askaryan emission
       from a cosmic ray-induced cascade in the lunar regolith, with a
       rough ice layer buried below the surface. \textit{b)} The
       time-domain electric field measured at a spacecraft \SI{25}{km}
       above the lunar surface from a similar simulation as
       \textit{a)}. The electric field is shown for pure ice (blue)
       and extremely impure ice (orange). \label{fig:fdtd-results}}
   \end{figure}

An annotated time snapshot of the direct and reflected electric fields
from a cosmic ray induced cascade in the lunar regolith with a
\SI{50}{cm} thick ice layer \SI{6}{m} below the surface is shown in
Figure~\ref{fig:fdtd-slice}; the corresponding time-domain electric
field waveform observed at a spacecraft orbiting \SI{25}{km} above the
surface is shown in Figure~\ref{fig:fdtd-waveforms}.

The first wavefronts (furthest to the upper right in blue and light
purple) are the \textit{direct} emission from the cascade, observed
without reflection, and are the first waveforms to arrive at the
spacecraft. As shown in Figure~\ref{fig:regolith-angular-spectrum} and
Figure~\ref{fig:ice-angular-spectrum}, as the observation direction
moves further off-cone the emitted power and total bandwidth
decreases, so the direct pulse observed at the spacecraft is typically
weaker and less impulsive than the more on-cone reflected pulses
(although, the direct pulses do not suffer from the attenuation due to
the longer path through the regolith of the reflected wavefronts or
the subsurface Fresnel reflection coefficient). This can been seen in the small amplitude and broad pulse width of the direct (orange) waveform in Figure~\ref{fig:fdtd-waveforms}. For events for which
the direct emission is identifiable, the relative timing between the
direct and reflected impulses provides a powerful fiducial for the
depth of each subsurface layer.

Since Askaryan radiation is strongly beamed around the Cherenkov angle
(\(\sim30^{\circ} - 60^{\circ}\) off the cosmic ray axis depending upon the
material---\({\sim}\SI{54}{^{\circ}}\) in the lunar regolith), the reflected
waveforms, seen as two bright green wavefronts in Figure~\ref{fig:fdtd-slice},
are significantly stronger and are highly impulsive. Even when including a rough
ice surface for the reflection, the corresponding pulse widths are only a few
nanoseconds in duration. In the geometry of this simulation, the regolith-ice
and ice-regolith reflections have inverted polarity with respect to each other
due to the change in sign of the Fresnel reflection coefficient. This clear
polarity inversion allows for distinguishing the pulses from the top and bottom
of any subsurface ice layer, and also acts as a direct probe of the thickness of
the subsurface reflector.


The inset in Figure~\ref{fig:fdtd-waveforms} shows the signal to noise ratio
achieved by this \SI{10}{EeV} cosmic ray as a function of the ice purity or
equivalently, the refractive index of the layer compared to the regolith. In
this case, at an orbital altitude of \SI{25}{km}, an Askaryan radar could
achieve an electric field signal to noise ratio (SNR) greater than two for ice
purities down to \({\sim}20\%\) (i.e. 80\% regolith by weight), allowing this technique to
explore a range of subsurface compositions.

Due to the computational cost of FDTD simulations, we are only able to
run a small number of discrete geometries, but they confirm the
validity of the fundamental principles of this technique.

\subsection{Monte Carlo Results}
\label{sec:monte-carlo-results}

\begin{figure}[htbp!]
     \centering

     \begin{subfigure}[b]{0.495\textwidth}
       \centering
       \includegraphics[width=\textwidth]{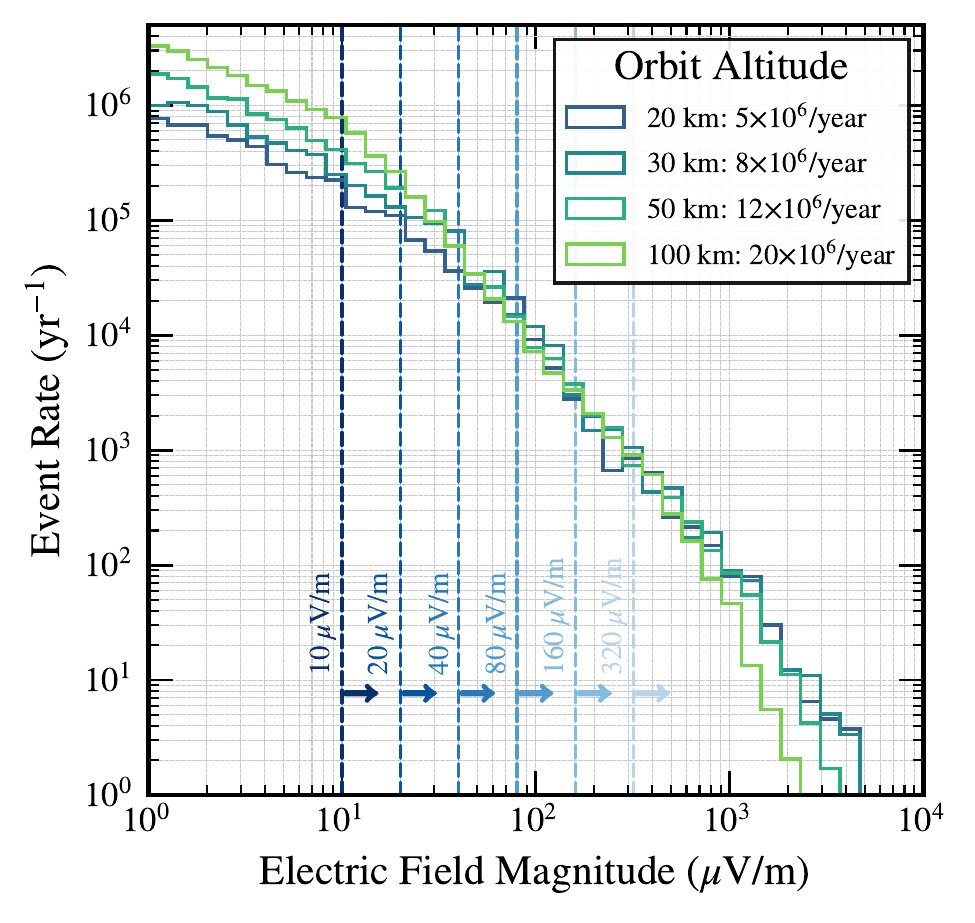}
       \caption{}
       \label{fig:event-efield-spectrum}
     \end{subfigure}%
     \hfill
     \begin{subfigure}[b]{0.495\textwidth}
       \centering
       \includegraphics[width=\textwidth]{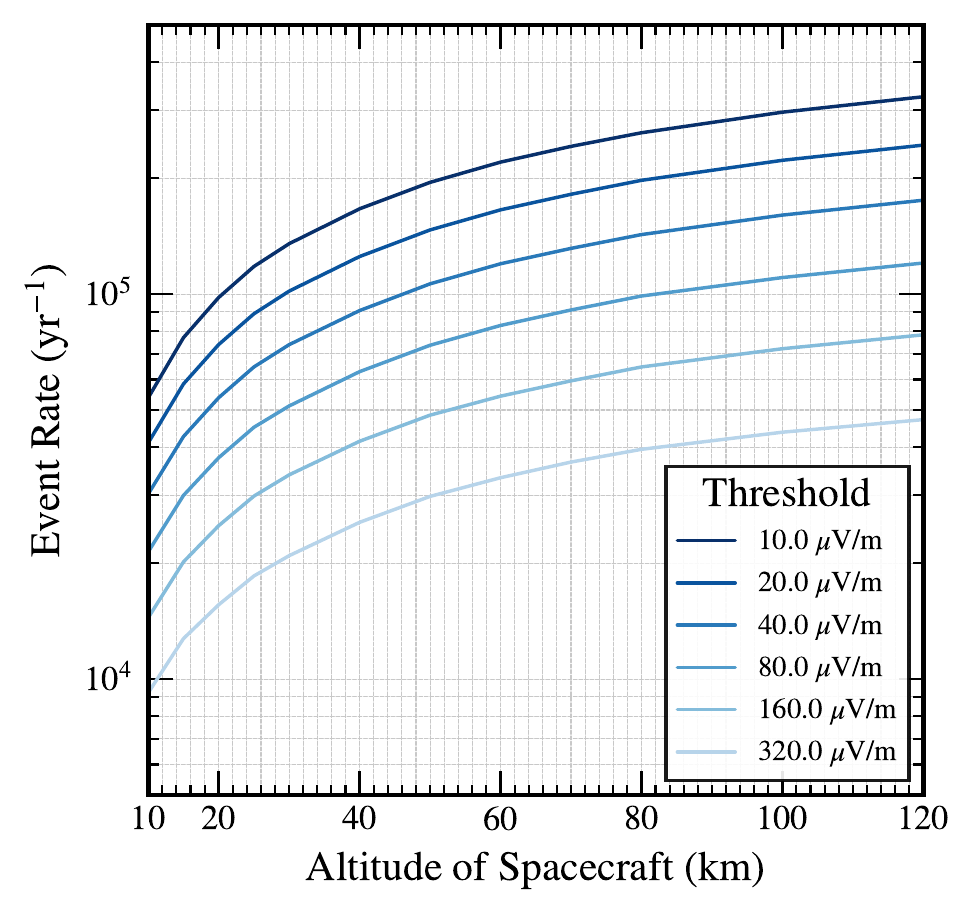}
       \caption{}
       \label{fig:event-altitude-scan}
     \end{subfigure}%

     \begin{subfigure}[b]{0.495\textwidth}
       \centering
       \includegraphics[width=\textwidth]{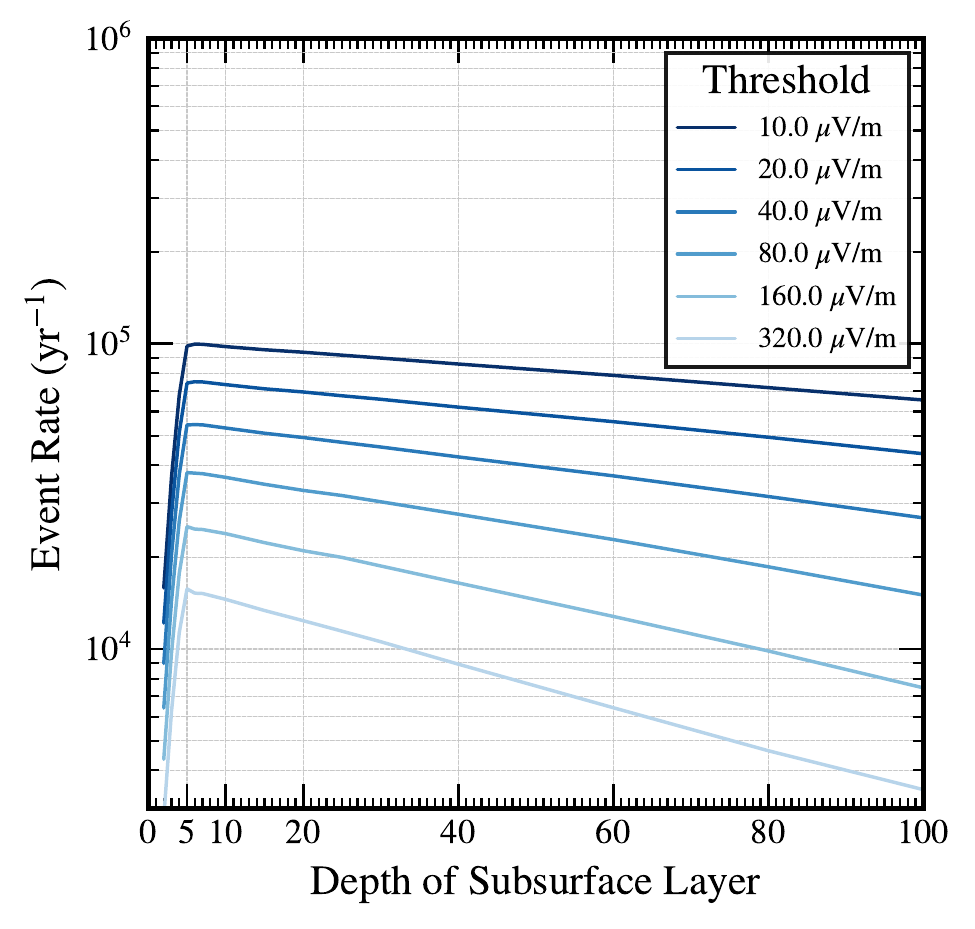}
       \caption{}
       \label{fig:event-depth-scan}
     \end{subfigure}
     \hfill
     \begin{subfigure}[b]{0.495\textwidth}
       \centering
       \includegraphics[width=\textwidth]{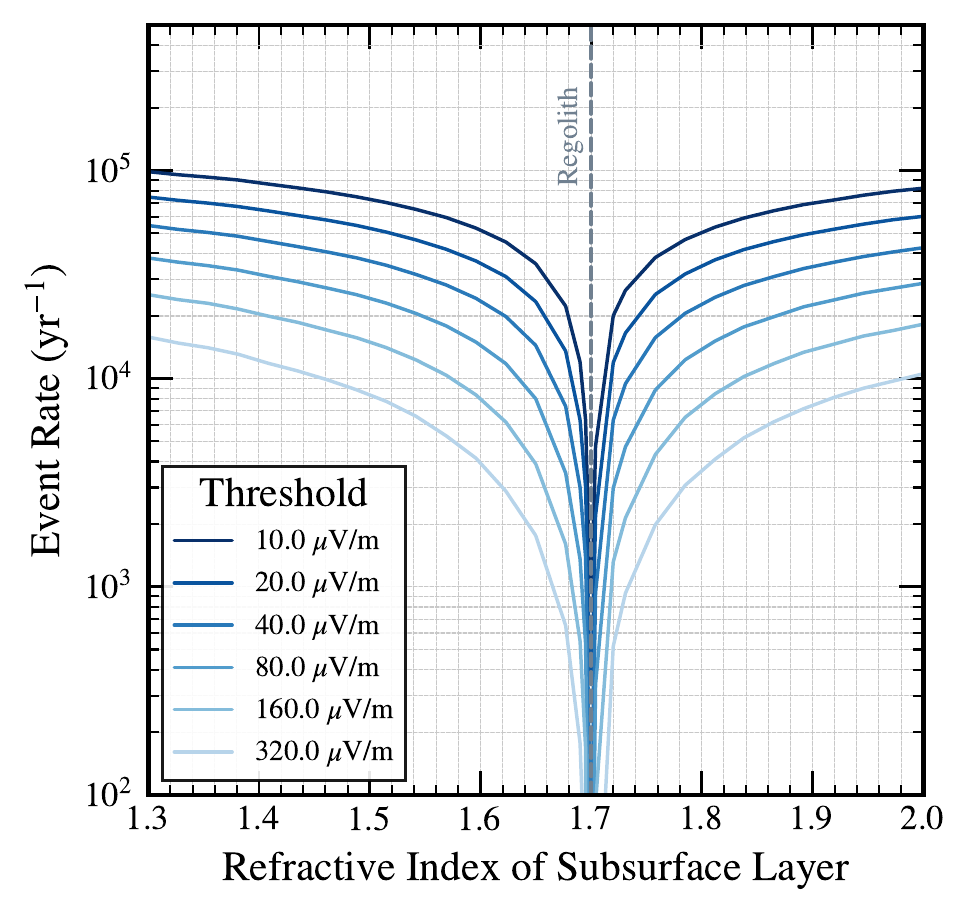}
       \caption{}
       \label{fig:event-index-scan}
     \end{subfigure}
     \caption{\textit{a):} The distribution of Askaryan electric field
       at the spacecraft from a uniform ice layer buried \SI{5}{m}
       below the lunar surface for various spacecraft orbital
       altitudes. The blue dashed vertical lines show the location of
       the trigger thresholds used in the other figures. \textit{b):}
       The detected event (\SI{}{yr^{-1}}) as a function of
       spacecraft altitude for various detection
       thresholds. \textit{c):} The detected event (\SI{}{yr^{-1}})
       as a function of the depth of the subsurface layer for various
       detection thresholds; these simulations include a radio
       attenuation length for the regolith which is responsible for
       the decline in event rate for deeper layers (see
       text). \textit{d):} The detected event (\SI{}{yr^{-1}}) as a
       function of the refractive index of the subsurface layer; the
       index of the regolith, where no dielectric contrast would be
       observed, is denoted by the vertical gray dashed line.~\label{fig:monte-carlo-results}}
   \end{figure}

The results from our Monte Carlo calculation of the subsurface detection rate
(Equation~\ref{eq:monte-carlo-evaluation}) for various subsurface and
instrumental parameters are shown in Figure~\ref{fig:monte-carlo-results}. Since
there is no active transmitter in an Askaryan radar, the overall event rate is
set by the combination of the sensitivity of the spacecraft, the geometry of the
detection, the (fixed) cosmic ray flux, and the (fixed) properties of the
subsurface (i.e. attenuation length, dielectric contrast with the volume, etc.).
Therefore, the geometry and spacecraft sensitivity are the two design parameters
available for increasing the rate of subsurface detections given a fixed subsurface composition, thus these are the primary parameters that we vary in our studies.

\subsubsection{Detectability of Askaryan Radar Signals}
\label{sec:discussion-detectability}

Since the radio emission detected by an Askaryan radar can come from anywhere
within the field of view of the spacecraft, at any time, with electric fields
extending orders of magnitude above and below the threshold of the instrument,
an Askaryan radar instrument must be able to continuously trigger on oncoming
radio signals that may \textit{potentially} be an Askaryan subsurface signal,
and localize the received radar emission back to a subsurface location. This
``trigger threshold'' is typically set so that the lowest signal-to-noise ratio
(SNR) subsurface reflections that could potentially be identified in an offline
analysis are captured. 

This can be accomplished with a multi-channel beamforming radar receiver whose
signals can be coherently combined in real-time to create
highly-directional synthetic \textit{beams} that can be used to
reconstruct the subsurface location of the received signal, even when
the single-antenna signal-to-noise ratio (SNR) is low. An
interferometric or multi-channel instrument design has extensive
heritage in the field of cosmic ray and neutrino astrophysics, where
it has been used by a host of existing
Askaryan detection experiments~\cite{2011_ARA,2019_ARIANNA,2006_ANITA_Intro,vieregg15:_techn_detec_pev_neutr_using}. In
particular, this was the technique used by the ANITA instrument to
identify its sample of ice-reflected cosmic rays during its
circumpolar orbit of Antarctica, with \({\sim}\SI{0.1}{^{\circ}}\) resolution
on the incoming radio direction, exactly as would be performed on a
potential future Askaryan radar
mission~\cite{2010_Hoover_CosmicRay_Discovery}.




\subsubsection{Maximum Subsurface Detection Rate}
\label{sec:discussion-event-rate}

There is a practically unlimited number of cosmic ray impacts on the surface of
airless bodies: \(\gtrsim35,000\)~per~\SI{}{km^{2}}~per~year above \SI{100}{PeV}
and \(\gtrsim\SI{e9}{}\)~per~\SI{}{km^{2}}~per~year above \SI{1}{PeV}, so the
detection rate is effectively determined by the minimum cosmic ray energy
detectable by the spacecraft (which is, in turn, determined by the electric
field threshold of the instrument) assuming there are subsurface reflectors
present. Figure~\ref{fig:event-efield-spectrum} shows the simulated distribution
of integrated electric field at the spacecraft from cosmic rays with
energies between \SI{500}{PeV} and \SI{500}{EeV}, for different (circular)
orbital altitudes of the spacecraft above the surface of the Moon. This energy
span is chosen as a realistic range that could plausibly be detected by a small
orbital Askaryan radar mission. Since the galactic cosmic ray flux extends for
\(\gtrsim9\)~orders of magnitude below this threshold, a larger, more sensitive,
mission could detect significantly more detections than presented here, with the
detection rate growing rapidly with decreasing threshold.

For each of these initial simulations, we assumed a nominal \SI{1}{m}
thick ice layer with a refractive index of 1.305 buried \SI{5}{m} below
the surface of the regolith. Due to the falling cosmic ray energy
spectrum, the electric field spectrum is also decreasing (since the
emitted Askaryan electric field strength scales linearly with cosmic
ray energy). The legend counts the total number of observed electric
fields (at the spacecraft) with total magnitude greater than \SI{1}{\mu V/m} when integrated over the frequency range between~\SI{50}{MHz}
and \SI{1}{GHz}. For the simulated lunar orbits, the total number of
subsurface reflections observed at the spacecraft exceeds
\SI{e6}{} in all cases, extending up to
\SI{2e7}{} for a \SI{100}{km} lunar
orbit. Figure~\ref{fig:event-efield-spectrum} also shows the six
different electric field thresholds, from \SI{10}{\mu V/m} to \SI{320}{\mu
  V/m}, that are used in the \textit{other} plots in
Figure~\ref{fig:monte-carlo-results} to demonstrate the realized
(detected) event rate for a spacecraft with the given electric field
sensitivity. The range of electric field thresholds were chosen to
cover a range of possible sensitivities that might be achievable with
a lunar orbiting Askaryan radar mission with a small beamformed antenna array.


The scaling of the total event rate as a function of orbit altitude is
shown in Figure~\ref{fig:event-altitude-scan}. A lower orbit restricts
the total geometric surface area viewable from the spacecraft (which
reduces the total event rate as cosmic rays impact uniformly in area),
but the reduced cosmic ray to spacecraft distance allows a greater
fraction of any reflections to be detected given a fixed spacecraft
sensitivity.

Alternatively, a higher orbit, with corresponding larger
surface area coverage, will view more cosmic ray impacts but with an
attenuated electric field strength, reducing the detection
efficiency. 
As shown in Figure~\ref{fig:event-altitude-scan}, an Askaryan radar is broadly
sensitive over a wide range of orbital altitudes, from \SI{10}{km} (a
potentially unstable orbit for the Moon) up to in excess of \SI{120}{km}, and
can be optimized for a specific application or performance (i.e. resolution on
the ground, requiring lower altitudes, or maximizing the number of events that
would require higher altitudes).

\subsubsection{Maximum Subsurface Detection Depth}
\label{sec:discussion-max-depth}

Unlike most other radar techniques, Askaryan subsurface radar does not have a
fixed depth limit but instead has a decreasing detection rate as the depth of
the subsurface reflector increases. This is shown in
Figure~\ref{fig:event-depth-scan}. As discussed in
Section~\ref{sec:askaryan-effect}, the power emitted via the Askaryan effect
scales quadratically with the primary energy of the cosmic ray, so the emitted
power will vary by \({\sim}6\) orders of magnitude for an instrument sensitive
to cosmic rays between \(\SI{0.5}{EeV}\) and \(\SI{500}{EeV}\). For near surface
reflectors, the total radio path length through the regolith is at a minimum, so
the majority of cosmic rays above the minimum energy are potentially detectable
and the event rate is maximized. As the subsurface layer under consideration
moves deeper, the regolith path length increases and a higher energy cosmic ray
is required in order to exceed the electric field threshold. Since the cosmic ray
energy spectrum is steeply falling, the higher energy cosmic rays that are
needed to probe progressively deeper reflectors are progressively more rare, 
decreasing the detection rate, but still allowing for the detection of ice layers down to depths in excess of \(\mathcal{O}(\SI{100}{m})\) with these rarer ultrahigh energy cosmic ray.

For very shallow reflectors (\({\lesssim}\SI{3}{m}\)), the cosmic ray induced
cascades have not completely developed and therefore emit less Askaryan power
(as there is an underdeveloped charge excess in the cascade). However, the
\textit{total} charge in the shower will emit coherent transition radiation
as it cross into, and out of, the subsurface layer that could also potentially be detected
by an Askaryan radar instrument, improving its sensitivity to very shallow reflectors; the detection and reconstruction of this
coherent radio transition radiation, and its use as a near-surface radar, will
be explored in a later work.






\section{Conclusion}
\label{sec:conclusion}

We have demonstrated preliminary simulations of a new passive radar
sounding technique, \textit{Askaryan subsurface radar}, that has the
potential to be a promising new method for performing subsurface radar
probes on airless or nearly airless planetary bodies. One of the prime
advantages of the Askaryan radar technique is that the radio emission
is generated beneath the surface, completely bypassing the effects of
surface clutter and dispersion that adversely affects traditional
orbital radar sounding. Our detailed FDTD and Monte Carlo simulations demonstrate that this technique would allow a small surface or orbital antenna array to detect buried subsurface ice layers in the regolith over the vast majority of the phase space predicted by current models, without any active radar instrumentation. This technique can be applied to \textit{any} subsurface body without a dense atmosphere, including icy satellites (such as Ganymede), and further work will adapt and improve these simulations for other surface penetrating planetary science applications.




\backmatter

\bibliographystyle{bst/sn-mathphys}
\bibliography{askaryan_radar}


\begin{thebibliography}{39}
\ifx \bisbn   \undefined \def \bisbn  #1{ISBN #1}\fi
\ifx \binits  \undefined \def \binits#1{#1}\fi
\ifx \bauthor  \undefined \def \bauthor#1{#1}\fi
\ifx \batitle  \undefined \def \batitle#1{#1}\fi
\ifx \bjtitle  \undefined \def \bjtitle#1{#1}\fi
\ifx \bvolume  \undefined \def \bvolume#1{\textbf{#1}}\fi
\ifx \byear  \undefined \def \byear#1{#1}\fi
\ifx \bissue  \undefined \def \bissue#1{#1}\fi
\ifx \bfpage  \undefined \def \bfpage#1{#1}\fi
\ifx \blpage  \undefined \def \blpage #1{#1}\fi
\ifx \burl  \undefined \def \burl#1{\textsf{#1}}\fi
\ifx \doiurl  \undefined \def \doiurl#1{\url{https://doi.org/#1}}\fi
\ifx \betal  \undefined \def \betal{\textit{et al.}}\fi
\ifx \binstitute  \undefined \def \binstitute#1{#1}\fi
\ifx \binstitutionaled  \undefined \def \binstitutionaled#1{#1}\fi
\ifx \bctitle  \undefined \def \bctitle#1{#1}\fi
\ifx \beditor  \undefined \def \beditor#1{#1}\fi
\ifx \bpublisher  \undefined \def \bpublisher#1{#1}\fi
\ifx \bbtitle  \undefined \def \bbtitle#1{#1}\fi
\ifx \bedition  \undefined \def \bedition#1{#1}\fi
\ifx \bseriesno  \undefined \def \bseriesno#1{#1}\fi
\ifx \blocation  \undefined \def \blocation#1{#1}\fi
\ifx \bsertitle  \undefined \def \bsertitle#1{#1}\fi
\ifx \bsnm \undefined \def \bsnm#1{#1}\fi
\ifx \bsuffix \undefined \def \bsuffix#1{#1}\fi
\ifx \bparticle \undefined \def \bparticle#1{#1}\fi
\ifx \barticle \undefined \def \barticle#1{#1}\fi
\bibcommenthead
\ifx \bconfdate \undefined \def \bconfdate #1{#1}\fi
\ifx \botherref \undefined \def \botherref #1{#1}\fi
\ifx \url \undefined \def \url#1{\textsf{#1}}\fi
\ifx \bchapter \undefined \def \bchapter#1{#1}\fi
\ifx \bbook \undefined \def \bbook#1{#1}\fi
\ifx \bcomment \undefined \def \bcomment#1{#1}\fi
\ifx \oauthor \undefined \def \oauthor#1{#1}\fi
\ifx \citeauthoryear \undefined \def \citeauthoryear#1{#1}\fi
\ifx \endbibitem  \undefined \def \endbibitem {}\fi
\ifx \bconflocation  \undefined \def \bconflocation#1{#1}\fi
\ifx \arxivurl  \undefined \def \arxivurl#1{\textsf{#1}}\fi
\csname PreBibitemsHook\endcsname

\bibitem{1974_Apollo_Lunar_Sounding_Radar}
\begin{barticle}
\bauthor{\bsnm{{Porcello}}, \binits{L.J.}},
\bauthor{\bsnm{{Jordan}}, \binits{R.L.}},
\bauthor{\bsnm{{Zelenka}}, \binits{J.S.}},
\bauthor{\bsnm{{Adams}}, \binits{G.F.}},
\bauthor{\bsnm{{Phillips}}, \binits{R.J.}},
\bauthor{\bsnm{{Brown}}, \binits{J.} \bsuffix{W.~E.}},
\bauthor{\bsnm{{Ward}}, \binits{S.H.}},
\bauthor{\bsnm{{Jackson}}, \binits{P.L.}}:
\batitle{{The Apollo lunar sounder radar system.}}
\bjtitle{IEEE Proceedings}
\bvolume{62},
\bfpage{769}--\blpage{783}
(\byear{1974})
\end{barticle}
\endbibitem

\bibitem{2000_LRS_Selene}
\begin{barticle}
\bauthor{\bsnm{Ono}, \binits{T.}},
\bauthor{\bsnm{Oya}, \binits{H.}}:
\batitle{Lunar radar sounder (lrs) experiment on-board the selene spacecraft}.
\bjtitle{Earth, Planets and Space}
\bvolume{52}(\bissue{9}),
\bfpage{629}--\blpage{637}
(\byear{2000}).
\doiurl{10.1186/bf03351671}
\end{barticle}
\endbibitem

\bibitem{2011_Insight_Lunar_Processes_MiniRF}
\begin{bchapter}
\bauthor{\bsnm{{Bussey}}, \binits{D.B.J.}},
\bauthor{\bsnm{{Spudis}}, \binits{P.D.}},
\bauthor{\bsnm{{Mini-Rf Team}}}:
\bctitle{{New Insights into Lunar Processes and History from Global Mapping by
  Mini-RF Radar}}.
In: \bbtitle{42nd Annual Lunar and Planetary Science Conference}.
\bsertitle{Lunar and Planetary Science Conference},
p. \bfpage{2086}
(\byear{2011})
\end{bchapter}
\endbibitem

\bibitem{2020_Change5_Lunar_Radar}
\begin{barticle}
\bauthor{\bsnm{Lu}, \binits{W.}},
\bauthor{\bsnm{Li}, \binits{Y.}},
\bauthor{\bsnm{Ji}, \binits{Y.}},
\bauthor{\bsnm{Tang}, \binits{C.}},
\bauthor{\bsnm{Zhou}, \binits{B.}},
\bauthor{\bsnm{Fang}, \binits{G.}}:
\batitle{Ultra-wideband mimo array for penetrating lunar regolith structures on
  the chang’e-5 lander}.
\bjtitle{Electronics}
\bvolume{10}(\bissue{1}),
\bfpage{8}
(\byear{2020}).
\doiurl{10.3390/electronics10010008}
\end{barticle}
\endbibitem

\bibitem{2007_SHARAD_MRO}
\begin{botherref}
\oauthor{\bsnm{Seu}, \binits{R.}},
\oauthor{\bsnm{Phillips}, \binits{R.J.}},
\oauthor{\bsnm{Biccari}, \binits{D.}},
\oauthor{\bsnm{Orosei}, \binits{R.}},
\oauthor{\bsnm{Masdea}, \binits{A.}},
\oauthor{\bsnm{Picardi}, \binits{G.}},
\oauthor{\bsnm{Safaeinili}, \binits{A.}},
\oauthor{\bsnm{Campbell}, \binits{B.A.}},
\oauthor{\bsnm{Plaut}, \binits{J.J.}},
\oauthor{\bsnm{Marinangeli}, \binits{L.}},
\oauthor{\bsnm{Smrekar}, \binits{S.E.}},
\oauthor{\bsnm{Nunes}, \binits{D.C.}}:
Sharad sounding radar on the mars reconnaissance orbiter.
Journal of Geophysical Research
\textbf{112}(E5)
(2007).
\doiurl{10.1029/2006je002745}
\end{botherref}
\endbibitem

\bibitem{2015_MARSIS_Radar}
\begin{barticle}
\bauthor{\bsnm{Orosei}, \binits{R.}},
\bauthor{\bsnm{Jordan}, \binits{R.L.}},
\bauthor{\bsnm{Morgan}, \binits{D.D.}},
\bauthor{\bsnm{Cartacci}, \binits{M.}},
\bauthor{\bsnm{Cicchetti}, \binits{A.}},
\bauthor{\bsnm{Duru}, \binits{F.}},
\bauthor{\bsnm{Gurnett}, \binits{D.A.}},
\bauthor{\bsnm{Heggy}, \binits{E.}},
\bauthor{\bsnm{Kirchner}, \binits{D.L.}},
\bauthor{\bsnm{Noschese}, \binits{R.}},
\bauthor{\bsnm{Kofman}, \binits{W.}},
\bauthor{\bsnm{Masdea}, \binits{A.}},
\bauthor{\bsnm{Plaut}, \binits{J.J.}},
\bauthor{\bsnm{Seu}, \binits{R.}},
\bauthor{\bsnm{Watters}, \binits{T.R.}},
\bauthor{\bsnm{Picardi}, \binits{G.}}:
\batitle{Mars advanced radar for subsurface and ionospheric sounding (marsis)
  after nine years of operation: A summary}.
\bjtitle{Planetary and Space Science}
\bvolume{112},
\bfpage{98}--\blpage{114}
(\byear{2015}).
\doiurl{10.1016/j.pss.2014.07.010}
\end{barticle}
\endbibitem

\bibitem{2020_RIMFAX_Mars}
\begin{botherref}
\oauthor{\bsnm{Hamran}, \binits{S.-E.}},
\oauthor{\bsnm{Paige}, \binits{D.A.}},
\oauthor{\bsnm{Amundsen}, \binits{H.E.F.}},
\oauthor{\bsnm{Berger}, \binits{T.}},
\oauthor{\bsnm{Brovoll}, \binits{S.}},
\oauthor{\bsnm{Carter}, \binits{L.}},
\oauthor{\bsnm{Damsgård}, \binits{L.}},
\oauthor{\bsnm{Dypvik}, \binits{H.}},
\oauthor{\bsnm{Eide}, \binits{J.}},
\oauthor{\bsnm{Eide}, \binits{S.}},
\oauthor{\bsnm{Ghent}, \binits{R.}},
\oauthor{\bsnm{Helleren}, \binits{O.}},
\oauthor{\bsnm{Kohler}, \binits{J.}},
\oauthor{\bsnm{Mellon}, \binits{M.}},
\oauthor{\bsnm{Nunes}, \binits{D.C.}},
\oauthor{\bsnm{Plettemeier}, \binits{D.}},
\oauthor{\bsnm{Rowe}, \binits{K.}},
\oauthor{\bsnm{Russell}, \binits{P.}},
\oauthor{\bsnm{Øyan}, \binits{M.J.}}:
Radar imager for mars’ subsurface experiment—rimfax.
Space Science Reviews
\textbf{216}(8)
(2020).
\doiurl{10.1007/s11214-020-00740-4}
\end{botherref}
\endbibitem

\bibitem{2015_REASON}
\begin{bchapter}
\bauthor{\bsnm{{Moussessian}}, \binits{A.}},
\bauthor{\bsnm{{Blankenship}}, \binits{D.D.}},
\bauthor{\bsnm{{Plaut}}, \binits{J.J.}},
\bauthor{\bsnm{{Patterson}}, \binits{G.W.}},
\bauthor{\bsnm{{Gim}}, \binits{Y.}},
\bauthor{\bsnm{{Schroeder}}, \binits{D.M.}},
\bauthor{\bsnm{{Soderlund}}, \binits{K.M.}},
\bauthor{\bsnm{{Grima}}, \binits{C.}},
\bauthor{\bsnm{{Young}}, \binits{D.A.}},
\bauthor{\bsnm{{Chapin}}, \binits{E.}}:
\bctitle{{REASON for Europa}}.
In: \bbtitle{AGU Fall Meeting Abstracts},
vol. \bseriesno{2015},
pp. \bfpage{13}--\blpage{05}
(\byear{2015})
\end{bchapter}
\endbibitem

\bibitem{2013_JUICE}
\begin{barticle}
\bauthor{\bsnm{{Grasset}}, \binits{O.}},
\bauthor{\bsnm{{Dougherty}}, \binits{M.K.}},
\bauthor{\bsnm{{Coustenis}}, \binits{A.}},
\bauthor{\bsnm{{Bunce}}, \binits{E.J.}},
\bauthor{\bsnm{{Erd}}, \binits{C.}},
\bauthor{\bsnm{{Titov}}, \binits{D.}},
\bauthor{\bsnm{{Blanc}}, \binits{M.}},
\bauthor{\bsnm{{Coates}}, \binits{A.}},
\bauthor{\bsnm{{Drossart}}, \binits{P.}},
\bauthor{\bsnm{{Fletcher}}, \binits{L.N.}},
\bauthor{\bsnm{{Hussmann}}, \binits{H.}},
\bauthor{\bsnm{{Jaumann}}, \binits{R.}},
\bauthor{\bsnm{{Krupp}}, \binits{N.}},
\bauthor{\bsnm{{Lebreton}}, \binits{J.-P.}},
\bauthor{\bsnm{{Prieto-Ballesteros}}, \binits{O.}},
\bauthor{\bsnm{{Tortora}}, \binits{P.}},
\bauthor{\bsnm{{Tosi}}, \binits{F.}},
\bauthor{\bsnm{{Van Hoolst}}, \binits{T.}}:
\batitle{{JUpiter ICy moons Explorer (JUICE): An ESA mission to orbit Ganymede
  and to characterise the Jupiter system}}.
\bjtitle{Planetary Space Science}
\bvolume{78},
\bfpage{1}--\blpage{21}
(\byear{2013}).
\doiurl{10.1016/j.pss.2012.12.002}
\end{barticle}
\endbibitem

\bibitem{2007_CONSERT_Comet_Radar}
\begin{barticle}
\bauthor{\bsnm{Kofman}, \binits{W.}},
\bauthor{\bsnm{Herique}, \binits{A.}},
\bauthor{\bsnm{Goutail}, \binits{J.-P.}},
\bauthor{\bsnm{Hagfors}, \binits{T.}},
\bauthor{\bsnm{Williams}, \binits{I.P.}},
\bauthor{\bsnm{Nielsen}, \binits{E.}},
\bauthor{\bsnm{Barriot}, \binits{J.-P.}},
\bauthor{\bsnm{Barbin}, \binits{Y.}},
\bauthor{\bsnm{Elachi}, \binits{C.}},
\bauthor{\bsnm{Edenhofer}, \binits{P.}},
\bauthor{\bsnm{Levasseur-Regourd}, \binits{A.-C.}},
\bauthor{\bsnm{Plettemeier}, \binits{D.}},
\bauthor{\bsnm{Picardi}, \binits{G.}},
\bauthor{\bsnm{Seu}, \binits{R.}},
\bauthor{\bsnm{Svedhem}, \binits{V.}}:
\batitle{The comet nucleus sounding experiment by radiowave transmission
  (consert): A short description of the instrument and of the commissioning
  stages}.
\bjtitle{Space Science Reviews}
\bvolume{128}(\bissue{1-4}),
\bfpage{413}--\blpage{432}
(\byear{2007}).
\doiurl{10.1007/s11214-006-9034-9}
\end{barticle}
\endbibitem

\bibitem{2016_Radar_SolarSystem_Review}
\begin{barticle}
\bauthor{\bsnm{Ciarletti}, \binits{V.}}:
\batitle{A variety of radars designed to explore the hidden structures and
  properties of the solar system’s planets and bodies}.
\bjtitle{Comptes Rendus Physique}
\bvolume{17}(\bissue{9}),
\bfpage{966}--\blpage{975}
(\byear{2016}).
\doiurl{10.1016/j.crhy.2016.07.022}
\end{barticle}
\endbibitem

\bibitem{2001_Knight_GPR_Environmental_Review}
\begin{barticle}
\bauthor{\bsnm{Knight}, \binits{R.}}:
\batitle{Ground penetrating radar for environmental applications}.
\bjtitle{Annual Review of Earth and Planetary Sciences}
\bvolume{29}(\bissue{1}),
\bfpage{229}--\blpage{255}
(\byear{2001}).
\doiurl{10.1146/annurev.earth.29.1.229}
\end{barticle}
\endbibitem

\bibitem{2015_Radar_Planetary_Science_Overview}
\begin{botherref}
\oauthor{\bsnm{Tosti}, \binits{F.}},
\oauthor{\bsnm{Pajewski}, \binits{L.}}:
Applications of radar systems in planetary sciences: An overview.
Springer Transactions in Civil and Environmental Engineering,
361--371
(2015).
\doiurl{10.1007/978-3-319-04813-0_15}
\end{botherref}
\endbibitem

\bibitem{2021_Costello_SecondaryImpactBurial_IceDepth}
\begin{botherref}
\oauthor{\bsnm{Costello}, \binits{E.S.}},
\oauthor{\bsnm{Ghent}, \binits{R.R.}},
\oauthor{\bsnm{Lucey}, \binits{P.G.}}:
Secondary impact burial and excavation gardening on the moon and the depth to
  ice in permanent shadow.
Journal of Geophysical Research: Planets
\textbf{126}(9)
(2021).
\doiurl{10.1029/2021je006933}
\end{botherref}
\endbibitem

\bibitem{PDG2021}
\begin{barticle}
\bauthor{\bsnm{Zyla}, \binits{P.A.}}, \betal:
\batitle{{Review of Particle Physics}}.
\bjtitle{PTEP}
\bvolume{2020}(\bissue{8}),
\bfpage{083}--\blpage{01}
(\byear{2020}).
\doiurl{10.1093/ptep/ptaa104}
\end{barticle}
\endbibitem

\bibitem{2010_Hoover_CosmicRay_Discovery}
\begin{barticle}
\bauthor{\bsnm{{Hoover}}, \binits{S.}},
\bauthor{\bsnm{{Nam}}, \binits{J.}},
\bauthor{\bsnm{{Gorham}}, \binits{P.W.}},
\bauthor{\bsnm{{Grashorn}}, \binits{E.}},
\bauthor{\bsnm{{Allison}}, \binits{P.}},
\bauthor{\bsnm{{Barwick}}, \binits{S.W.}},
\bauthor{\bsnm{{Beatty}}, \binits{J.J.}},
\bauthor{\bsnm{{Belov}}, \binits{K.}},
\bauthor{\bsnm{{Besson}}, \binits{D.Z.}},
\bauthor{\bsnm{{Binns}}, \binits{W.R.}},
\bauthor{\bsnm{{Chen}}, \binits{C.}},
\bauthor{\bsnm{{Chen}}, \binits{P.}},
\bauthor{\bsnm{{Clem}}, \binits{J.M.}},
\bauthor{\bsnm{{Connolly}}, \binits{A.}},
\bauthor{\bsnm{{Dowkontt}}, \binits{P.F.}},
\bauthor{\bsnm{{Duvernois}}, \binits{M.A.}},
\bauthor{\bsnm{{Field}}, \binits{R.C.}},
\bauthor{\bsnm{{Goldstein}}, \binits{D.}},
\bauthor{\bsnm{{Vieregg}}, \binits{A.G.}},
\bauthor{\bsnm{{Hast}}, \binits{C.}},
\bauthor{\bsnm{{Israel}}, \binits{M.H.}},
\bauthor{\bsnm{{Javaid}}, \binits{A.}},
\bauthor{\bsnm{{Kowalski}}, \binits{J.}},
\bauthor{\bsnm{{Learned}}, \binits{J.G.}},
\bauthor{\bsnm{{Liewer}}, \binits{K.M.}},
\bauthor{\bsnm{{Link}}, \binits{J.T.}},
\bauthor{\bsnm{{Lusczek}}, \binits{E.}},
\bauthor{\bsnm{{Matsuno}}, \binits{S.}},
\bauthor{\bsnm{{Mercurio}}, \binits{B.C.}},
\bauthor{\bsnm{{Miki}}, \binits{C.}},
\bauthor{\bsnm{{Mio{\v{c}}inovi{\'c}}}, \binits{P.}},
\bauthor{\bsnm{{Naudet}}, \binits{C.J.}},
\bauthor{\bsnm{{Ng}}, \binits{J.}},
\bauthor{\bsnm{{Nichol}}, \binits{R.J.}},
\bauthor{\bsnm{{Palladino}}, \binits{K.}},
\bauthor{\bsnm{{Reil}}, \binits{K.}},
\bauthor{\bsnm{{Romero-Wolf}}, \binits{A.}},
\bauthor{\bsnm{{Rosen}}, \binits{M.}},
\bauthor{\bsnm{{Ruckman}}, \binits{L.}},
\bauthor{\bsnm{{Saltzberg}}, \binits{D.}},
\bauthor{\bsnm{{Seckel}}, \binits{D.}},
\bauthor{\bsnm{{Varner}}, \binits{G.S.}},
\bauthor{\bsnm{{Walz}}, \binits{D.}},
\bauthor{\bsnm{{Wu}}, \binits{F.}}:
\batitle{{Observation of Ultrahigh-Energy Cosmic Rays with the ANITA
  Balloon-Borne Radio Interferometer}}.
\bjtitle{Physical Review Letters}
\bvolume{105}(\bissue{15}),
\bfpage{151101}
(\byear{2010})
{\href{https://arxiv.org/abs/1005.0035}{{arXiv:1005.0035}}}
{[astro-ph.HE]}.
\doiurl{10.1103/PhysRevLett.105.151101}
\end{barticle}
\endbibitem

\bibitem{2017_Auger_Combined_Fit}
\begin{barticle}
\bauthor{\bsnm{Aab}, \binits{A.}},
\bauthor{\bsnm{Abreu}, \binits{P.}},
\bauthor{\bsnm{Aglietta}, \binits{M.}},
\bauthor{\bsnm{Samarai}, \binits{I.A.}},
\bauthor{\bsnm{Albuquerque}, \binits{I.F.M.}},
\bauthor{\bsnm{Allekotte}, \binits{I.}},
\bauthor{\bsnm{Almela}, \binits{A.}},
\bauthor{\bsnm{Castillo}, \binits{J.A.}},
\bauthor{\bsnm{Alvarez-Mu\~niz}, \binits{J.}},
\bauthor{\bsnm{Anastasi}, \binits{G.A.}},
\bauthor{\bparticle{et} \bsnm{al.}}:
\batitle{Combined fit of spectrum and composition data as measured by the
  pierre auger observatory}.
\bjtitle{Journal of Cosmology and Astroparticle Physics}
\bvolume{2017}(\bissue{04}),
\bfpage{038}--\blpage{038}
(\byear{2017}).
\doiurl{10.1088/1475-7516/2017/04/038}
\end{barticle}
\endbibitem

\bibitem{2009_Antarctic_Ice_Sheet}
\begin{barticle}
\bauthor{\bsnm{Rémy}, \binits{F.}},
\bauthor{\bsnm{Parouty}, \binits{S.}}:
\batitle{Antarctic ice sheet and radar altimetry: A review}.
\bjtitle{Remote Sensing}
\bvolume{1}(\bissue{4}),
\bfpage{1212}--\blpage{1239}
(\byear{2009}).
\doiurl{10.3390/rs1041212}
\end{barticle}
\endbibitem

\bibitem{2017_Verzi_EnergySpectrum_UHECR}
\begin{botherref}
\oauthor{\bsnm{Verzi}, \binits{V.}},
\oauthor{\bsnm{Ivanov}, \binits{D.}},
\oauthor{\bsnm{Tsunesada}, \binits{Y.}}:
Measurement of energy spectrum of ultra-high energy cosmic rays.
Progress of Theoretical and Experimental Physics
\textbf{2017}(12)
(2017).
\doiurl{10.1093/ptep/ptx082}
\end{botherref}
\endbibitem

\bibitem{2010_TIERRAS}
\begin{barticle}
\bauthor{\bsnm{Tueros}, \binits{M.}},
\bauthor{\bsnm{Sciutto}, \binits{S.J.}}:
\batitle{Tierras: A package to simulate high energy cosmic ray showers
  underground, underwater and under-ice}.
\bjtitle{Comput. Phys. Commun.}
\bvolume{181},
\bfpage{380}--\blpage{392}
(\byear{2010})
\end{barticle}
\endbibitem

\bibitem{2012_MonteCarlo_AtmosphericShowers_ZHAireS}
\begin{barticle}
\bauthor{\bsnm{{Alvarez-Mu{\~n}iz}}, \binits{J.}},
\bauthor{\bsnm{{Carvalho}}, \binits{W.R.}},
\bauthor{\bsnm{{Zas}}, \binits{E.}}:
\batitle{{Monte Carlo simulations of radio pulses in atmospheric showers using
  ZHAireS}}.
\bjtitle{Astroparticle Physics}
\bvolume{35}(\bissue{6}),
\bfpage{325}--\blpage{341}
(\byear{2012})
{\href{https://arxiv.org/abs/1107.1189}{{arXiv:1107.1189}}}
{[astro-ph.HE]}.
\doiurl{10.1016/j.astropartphys.2011.10.005}
\end{barticle}
\endbibitem

\bibitem{1960_Askaryan_Prediction}
\begin{barticle}
\bauthor{\bsnm{Askar'yan}, \binits{G.A.}}:
\batitle{{Excess negative charge of an electron-photon shower and its coherent
  radio emission}}.
\bjtitle{Zh. Eksp. Teor. Fiz.}
\bvolume{41},
\bfpage{616}--\blpage{618}
(\byear{1961})
\end{barticle}
\endbibitem

\bibitem{Cherenkov1934}
\begin{botherref}
\oauthor{\bsnm{Cherenkov}, \binits{P.A.}}:
Visible emission of clean liquids by action of $\gamma$ radiation.
Doklady Akademii Nuak SSSR
\textbf{451}
(1934)
\end{botherref}
\endbibitem

\bibitem{1960_Tamm_CherenkovRadiation}
\begin{barticle}
\bauthor{\bsnm{Tamm}, \binits{I.E.}}:
\batitle{General characteristics of vavilov-cherenkov radiation}.
\bjtitle{Science}
\bvolume{131}(\bissue{3395}),
\bfpage{206}--\blpage{210}
(\byear{1960}).
\doiurl{10.1126/science.131.3395.206}
\end{barticle}
\endbibitem

\bibitem{Saltzberg:2000bk}
\begin{barticle}
\bauthor{\bsnm{Saltzberg}, \binits{D.}},
\bauthor{\bsnm{Gorham}, \binits{P.}},
\bauthor{\bsnm{Walz}, \binits{D.}},
\bauthor{\bsnm{Field}, \binits{C.}},
\bauthor{\bsnm{Iverson}, \binits{R.}},
\bauthor{\bsnm{Odian}, \binits{A.}},
\bauthor{\bsnm{Resch}, \binits{G.}},
\bauthor{\bsnm{Schoessow}, \binits{P.}},
\bauthor{\bsnm{Williams}, \binits{D.}}:
\batitle{{Observation of the Askaryan effect: Coherent microwave Cherenkov
  emission from charge asymmetry in high-energy particle cascades}}.
\bjtitle{Phys. Rev. Lett.}
\bvolume{86},
\bfpage{2802}--\blpage{2805}
(\byear{2001})
{\href{https://arxiv.org/abs/hep-ex/0011001}{{arXiv:hep-ex/0011001}}}.
\doiurl{10.1103/PhysRevLett.86.2802}
\end{barticle}
\endbibitem

\bibitem{Gorham:2004ny}
\begin{barticle}
\bauthor{\bsnm{Gorham}, \binits{P.W.}},
\bauthor{\bsnm{Saltzberg}, \binits{D.}},
\bauthor{\bsnm{Field}, \binits{R.C.}},
\bauthor{\bsnm{Guillian}, \binits{E.}},
\bauthor{\bsnm{Milincic}, \binits{R.}},
\bauthor{\bsnm{Walz}, \binits{D.}},
\bauthor{\bsnm{Williams}, \binits{D.}}:
\batitle{{Accelerator measurements of the Askaryan effect in rock salt: A
  Roadmap toward teraton underground neutrino detectors}}.
\bjtitle{Phys. Rev. D}
\bvolume{72},
\bfpage{023002}
(\byear{2005})
{\href{https://arxiv.org/abs/astro-ph/0412128}{{arXiv:astro-ph/0412128}}}.
\doiurl{10.1103/PhysRevD.72.023002}
\end{barticle}
\endbibitem

\bibitem{ANITA:2006nif}
\begin{barticle}
\bauthor{\bsnm{Gorham}, \binits{P.W.}}, \betal:
\batitle{{Observations of the Askaryan effect in ice}}.
\bjtitle{Phys. Rev. Lett.}
\bvolume{99},
\bfpage{171101}
(\byear{2007})
{\href{https://arxiv.org/abs/hep-ex/0611008}{{arXiv:hep-ex/0611008}}}.
\doiurl{10.1103/PhysRevLett.99.171101}
\end{barticle}
\endbibitem

\bibitem{Gorham:2017nzv}
\begin{barticle}
\bauthor{\bsnm{Gorham}, \binits{P.W.}}, \betal:
\batitle{{Picosecond timing of Microwave Cherenkov Impulses from High-Energy
  Particle Showers Using Dielectric-loaded Waveguides}}.
\bjtitle{Phys. Rev. Accel. Beams}
\bvolume{21}(\bissue{7}),
\bfpage{072901}
(\byear{2018})
{\href{https://arxiv.org/abs/1708.01798}{{arXiv:1708.01798}}}
{[physics.ins-det]}.
\doiurl{10.1103/PhysRevAccelBeams.21.072901}
\end{barticle}
\endbibitem

\bibitem{2011_ARA}
\begin{botherref}
\oauthor{\bsnm{Allison}, \binits{P.}},
\oauthor{\bsnm{Auffenberg}, \binits{J.}},
\oauthor{\bsnm{Bard}, \binits{R.}},
\oauthor{\bsnm{Beatty}, \binits{J.J.}},
\oauthor{\bsnm{Besson}, \binits{D.Z.}},
\oauthor{\bsnm{Boeser}, \binits{S.}},
\oauthor{\bsnm{Chen}, \binits{C.}},
\oauthor{\bsnm{Chen}, \binits{P.}},
\oauthor{\bsnm{Connolly}, \binits{A.}},
\oauthor{\bsnm{Davies}, \binits{J.}},
\oauthor{\bsnm{DuVernois}, \binits{M.}},
\oauthor{\bsnm{Fox}, \binits{B.}},
\oauthor{\bsnm{Gorham}, \binits{P.W.}},
\oauthor{\bsnm{Grashorn}, \binits{E.W.}},
\oauthor{\bsnm{Hanson}, \binits{K.}},
\oauthor{\bsnm{Haugen}, \binits{J.}},
\oauthor{\bsnm{Helbing}, \binits{K.}},
\oauthor{\bsnm{Hill}, \binits{B.}},
\oauthor{\bsnm{Hoffman}, \binits{K.D.}},
\oauthor{\bsnm{Huang}, \binits{M.}},
\oauthor{\bsnm{Huang}, \binits{M.H.A.}},
\oauthor{\bsnm{Ishihara}, \binits{A.}},
\oauthor{\bsnm{Karle}, \binits{A.}},
\oauthor{\bsnm{Kennedy}, \binits{D.}},
\oauthor{\bsnm{Landsman}, \binits{H.}},
\oauthor{\bsnm{Laundrie}, \binits{A.}},
\oauthor{\bsnm{Liu}, \binits{T.C.}},
\oauthor{\bsnm{Macchiarulo}, \binits{L.}},
\oauthor{\bsnm{Mase}, \binits{K.}},
\oauthor{\bsnm{Meures}, \binits{T.}},
\oauthor{\bsnm{Meyhandan}, \binits{R.}},
\oauthor{\bsnm{Miki}, \binits{C.}},
\oauthor{\bsnm{Morse}, \binits{R.}},
\oauthor{\bsnm{Newcomb}, \binits{M.}},
\oauthor{\bsnm{Nichol}, \binits{R.J.}},
\oauthor{\bsnm{Ratzlaff}, \binits{K.}},
\oauthor{\bsnm{Richman}, \binits{M.}},
\oauthor{\bsnm{Ritter}, \binits{L.}},
\oauthor{\bsnm{Rotter}, \binits{B.}},
\oauthor{\bsnm{Sandstrom}, \binits{P.}},
\oauthor{\bsnm{Seckel}, \binits{D.}},
\oauthor{\bsnm{Touart}, \binits{J.}},
\oauthor{\bsnm{Varner}, \binits{G.S.}},
\oauthor{\bsnm{Wang}, \binits{M.-Z.}},
\oauthor{\bsnm{Weaver}, \binits{C.}},
\oauthor{\bsnm{Wendorff}, \binits{A.}},
\oauthor{\bsnm{Yoshida}, \binits{S.}},
\oauthor{\bsnm{Young}, \binits{R.}}:
Design and initial performance of the askaryan radio array prototype eev
  neutrino detector at the south pole
(2011)
{\href{https://arxiv.org/abs/1105.2854v2}{{arXiv:1105.2854v2}}}.
\doiurl{10.1016/j.icarus.2012.05.028}
\end{botherref}
\endbibitem

\bibitem{2019_ARIANNA}
\begin{barticle}
\bauthor{\bsnm{Anker}, \binits{A.}},
\bauthor{\bsnm{Barwick}, \binits{S.W.}},
\bauthor{\bsnm{Bernhoff}, \binits{H.}},
\bauthor{\bsnm{Besson}, \binits{D.Z.}},
\bauthor{\bsnm{Bingefors}, \binits{N.}},
\bauthor{\bsnm{Gaswint}, \binits{G.}},
\bauthor{\bsnm{Glaser}, \binits{C.}},
\bauthor{\bsnm{Hallgren}, \binits{A.}},
\bauthor{\bsnm{Hanson}, \binits{J.C.}},
\bauthor{\bsnm{Lahmann}, \binits{R.}},
\bauthor{\bsnm{Latif}, \binits{U.}},
\bauthor{\bsnm{Nam}, \binits{J.}},
\bauthor{\bsnm{Novikov}, \binits{A.}},
\bauthor{\bsnm{Klein}, \binits{S.R.}},
\bauthor{\bsnm{Kleinfelder}, \binits{S.A.}},
\bauthor{\bsnm{Nelles}, \binits{A.}},
\bauthor{\bsnm{Paul}, \binits{M.P.}},
\bauthor{\bsnm{Persichilli}, \binits{C.}},
\bauthor{\bsnm{Shively}, \binits{S.R.}},
\bauthor{\bsnm{Tatar}, \binits{J.}},
\bauthor{\bsnm{Unger}, \binits{E.}},
\bauthor{\bsnm{Wang}, \binits{S.-H.}},
\bauthor{\bsnm{Yodh}, \binits{G.}}:
\batitle{Targeting ultra-high energy neutrinos with the arianna experiment}.
\bjtitle{Advances in Space Research 64 (2019) 2595-2609}
(\byear{2019})
{\href{https://arxiv.org/abs/1903.01609v2}{{arXiv:1903.01609v2}}}.
\doiurl{10.1016/j.asr.2019.06.016}
\end{barticle}
\endbibitem

\bibitem{2006_ANITA_Intro}
\begin{barticle}
\bauthor{\bsnm{Gorham}, \binits{P.W.}}:
\batitle{The {ANITA} cosmogenic neutrino experiment}.
\bjtitle{Acoustic and Radio EeV Neutrino Detection Activities}
(\byear{2006}).
\doiurl{10.1142/9789812773791_0029}
\end{barticle}
\endbibitem

\bibitem{2011_alvarez-muniz:practical_askaryan}
\begin{barticle}
\bauthor{\bsnm{Alvarez-Mu\~{n}iz}, \binits{J.}},
\bauthor{\bsnm{Romero-Wolf}, \binits{A.}},
\bauthor{\bsnm{Zas}, \binits{E.}}:
\batitle{Practical and accurate calculations of askaryan radiation}.
\bjtitle{Phys. Rev. D 84, 103003 (2011)}
(\byear{2011})
{\href{https://arxiv.org/abs/1106.6283v3}{{arXiv:1106.6283v3}}}.
\doiurl{10.1103/PhysRevD.84.103003}
\end{barticle}
\endbibitem

\bibitem{1966_FDTD_Yee}
\begin{barticle}
\bauthor{\bsnm{Yee}, \binits{K.}}:
\batitle{Numerical solution of initial boundary value problems involving
  maxwell's equations in isotropic media}.
\bjtitle{IEEE Transactions on Antennas and Propagation}
\bvolume{14}(\bissue{3}),
\bfpage{302}--\blpage{307}
(\byear{1966}).
\doiurl{10.1109/TAP.1966.1138693}
\end{barticle}
\endbibitem

\bibitem{xfdtd}
\begin{botherref}
\oauthor{\bsnm{{Remcom Incorporated}}}:
XFdtd.
\url{https://www.remcom.com/xfdtd-3d-em-simulation-software}
\end{botherref}
\endbibitem

\bibitem{2014_LPR_Change3}
\begin{barticle}
\bauthor{\bsnm{Fang}, \binits{G.-Y.}},
\bauthor{\bsnm{Zhou}, \binits{B.}},
\bauthor{\bsnm{Ji}, \binits{Y.-C.}},
\bauthor{\bsnm{Zhang}, \binits{Q.-Y.}},
\bauthor{\bsnm{Shen}, \binits{S.-X.}},
\bauthor{\bsnm{Li}, \binits{Y.-X.}},
\bauthor{\bsnm{Guan}, \binits{H.-F.}},
\bauthor{\bsnm{Tang}, \binits{C.-J.}},
\bauthor{\bsnm{Gao}, \binits{Y.-Z.}},
\bauthor{\bsnm{Lu}, \binits{W.}},
\bauthor{\bsnm{Ye}, \binits{S.-B.}},
\bauthor{\bsnm{Han}, \binits{H.-D.}},
\bauthor{\bsnm{Zheng}, \binits{J.}},
\bauthor{\bsnm{Wang}, \binits{S.-Z.}}:
\batitle{Lunar penetrating radar onboard the chang’e-3 mission}.
\bjtitle{Research in Astronomy and Astrophysics}
\bvolume{14}(\bissue{12}),
\bfpage{1607}--\blpage{1622}
(\byear{2014}).
\doiurl{10.1088/1674-4527/14/12/009}
\end{barticle}
\endbibitem

\bibitem{2017_3D_Geologic_Model_Change3}
\begin{barticle}
\bauthor{\bsnm{Yuan}, \binits{Y.}},
\bauthor{\bsnm{Zhu}, \binits{P.}},
\bauthor{\bsnm{Zhao}, \binits{N.}},
\bauthor{\bsnm{Xiao}, \binits{L.}},
\bauthor{\bsnm{Garnero}, \binits{E.}},
\bauthor{\bsnm{Xiao}, \binits{Z.}},
\bauthor{\bsnm{Zhao}, \binits{J.}},
\bauthor{\bsnm{Qiao}, \binits{L.}}:
\batitle{The 3-d geological model around chang’e-3 landing site based on
  lunar penetrating radar channel 1 data}.
\bjtitle{Geophysical Research Letters}
\bvolume{44}(\bissue{13}),
\bfpage{6553}--\blpage{6561}
(\byear{2017}).
\doiurl{10.1002/2017gl073589}
\end{barticle}
\endbibitem

\bibitem{Hurst}
\begin{barticle}
\bauthor{\bsnm{{Shepard}}, \binits{M.K.}},
\bauthor{\bsnm{{Brackett}}, \binits{R.A.}},
\bauthor{\bsnm{{Arvidson}}, \binits{R.E.}}:
\batitle{{Self-affine (fractal) topography: Surface parameterization and radar
  scattering}}.
\bjtitle{Journal of Geophysical Research}
\bvolume{100}(\bissue{E6}),
\bfpage{11709}--\blpage{11718}
(\byear{1995}).
\doiurl{10.1029/95JE00664}
\end{barticle}
\endbibitem

\bibitem{2017_Combined_Fit_Spectrum_Composition}
\begin{barticle}
\bauthor{\bsnm{Aab}, \binits{A.}},
\bauthor{\bsnm{Abreu}, \binits{P.}},
\bauthor{\bsnm{Aglietta}, \binits{M.}},
\bauthor{\bsnm{Samarai}, \binits{I.A.}},
\bauthor{\bsnm{Albuquerque}, \binits{I.F.M.}},
\bauthor{\bsnm{Allekotte}, \binits{I.}},
\bauthor{\bsnm{Almela}, \binits{A.}},
\bauthor{\bsnm{Castillo}, \binits{J.A.}},
\bauthor{\bsnm{Alvarez-Mu\~niz}, \binits{J.}},
\bauthor{\bsnm{Anastasi}, \binits{G.A.}},
\bauthor{\bparticle{et} \bsnm{al.}}:
\batitle{Combined fit of spectrum and composition data as measured by the
  pierre auger observatory}.
\bjtitle{Journal of Cosmology and Astroparticle Physics}
\bvolume{2017}(\bissue{04}),
\bfpage{038}--\blpage{038}
(\byear{2017}).
\doiurl{10.1088/1475-7516/2017/04/038}
\end{barticle}
\endbibitem

\bibitem{vieregg15:_techn_detec_pev_neutr_using}
\begin{barticle}
\bauthor{\bsnm{Vieregg}, \binits{A.G.}},
\bauthor{\bsnm{Bechtol}, \binits{K.}},
\bauthor{\bsnm{Romero-Wolf}, \binits{A.}}:
\batitle{A technique for detection of pev neutrinos using a phased radio
  array}.
\bjtitle{JCAP 2 (2016) 005}
(\byear{2015})
{\href{https://arxiv.org/abs/1504.08006v2}{{arXiv:1504.08006v2}}}.
\doiurl{10.1088/1475-7516/2016/02/005}
\end{barticle}
\endbibitem

\end{thebibliography}


\end{document}